\documentclass[12pt]{article}
\usepackage{amsmath,amsfonts,amssymb,setspace}
\usepackage[utf8]{inputenc}
\usepackage[nosort]{cite}
\usepackage[usenames,table]{xcolor}
\usepackage{dsfont}
\usepackage{mathrsfs}
\usepackage{graphicx}	% for figures
\usepackage{float}
\usepackage{longtable}	% for breaking the table on multiple pages
\usepackage{lipsum}
\usepackage{verbatim}
\usepackage{leftidx}	% for indices
\usepackage{bm}			% mathbold use: \bm{expression}
\usepackage{marginnote}	% margin comments

\numberwithin{equation}{section} % eqs. numbering section

\usepackage[top=2.8cm,
			bottom=2.8cm,
			left=2.0cm,
			right=2.0cm]{geometry}
\usepackage[colorlinks=true,
			linkcolor=black,
			citecolor=black,
			urlcolor=blue,
			filecolor=black]{hyperref}

\allowdisplaybreaks
\begin{document}

\vspace{30pt}

\begin{center}

%%------------ TITLE -----------------------------------------------

{\Large\sc Fractional Particle and Sigma Model \\[12pt]}

%\vskip 1cm
\vspace{-5pt}
\par\noindent\rule{457pt}{0.5pt}

%%------------ AUTHORS --------------------------------------------

\vskip 1cm

{\sc Ion V. Vancea }

%%----------- ADDRESSES -------------------------------------------

\vspace{10pt}
{\it %\small
Group of Theoretical Physics and Mathematical Physics,\\
Department of Physics, Federal Rural University of Rio de Janeiro,\\
Cx. Postal 23851, BR 465 Km 7, 23890-000 Serop\'{e}dica - RJ,
Brazil}

\vspace{4pt}

%%----------- EMAILS ----------------------------------------------

{\tt\small 
\href{mailto:ionvancea@ufrrj.br}{ionvancea@ufrrj.br}
}

%------------ ABSTRACT -------------------------------------------

\vspace{30pt} {\sc\large Abstract} \end{center}

\noindent

We introduce a classical fractional particle model in $\mathbb{R}^{n}$, extending the Newtonian particle concept with the incorporation of the fractional Laplacian. A comprehensive discussion on kinetic properties, including linear momentum and kinetic energy, is provided. We further derive the equations of motion and discuss the symmetries of the system. The Green's function method is employed to solve the equations of motion in a general case. We illustrate the theory with three important examples: the free fractional particle, the fractional harmonic oscillator, and the charged fractional particle that interacts locally with the electromagnetic field.  We use the results of the extension problem by Caffarelli and Silvestre, to construct the associated classical local sigma model for the fractional particle. The sigma model is then quantized using the canonical quantization method, and we compute the vacuum energy at the boundary.

%------------ END TITLE PAGE -------------------------------------

%------------ PAPER BODY -----------------------------------------

\newpage

%-----------------------------------------------------------------

\section{Introduction}

Motivated by the consideration of non-local effects within the framework of string cosmology, new non-local field models emerged through a reformulation of initial conditions employing the diffusion method, as proposed in \cite{Calcagni:2007ru,Calcagni:2007ef}, with further developments discussed in \cite{Calcagni:2018lyd}. This approach draws inspiration from the bosonic cubic string field theory elucidated by \cite{Witten:1985cc}. As observed in \cite{Mintchev:2001yz}, reveals that by imposing field commutativity solely on the brane, local fields on the brane induce non-local fields in both Minkowski and anti-de Sitter spacetimes.

However, distinct from such scenarios, wherein local fields induce non-local effects, the specific form of non-locality generated by fractional Laplacians provides a reverse mapping. Remarkably, it is demonstrated that the non-local scalar field on the brane can be transformed into a local field theory in the bulk through the application of the extension problem formulated by Caffarelli and Silvestre \cite{Caffarelli:2007}. This transformative mapping has been recently employed to investigate the conformal invariance of fractional Lagrangian field theories \cite{Rajabpour:2011qr} and the long-range Ising model \cite{Paulos:2015jfa}, yielding results consistent with the AdS/CFT method. Additionally, the canonical quantization of the local scalar theory in the bulk has been recently addressed in \cite{Frassino:2019yip}. 
An alternative methodology for quantizing field theories has been previously delineated, based on the power series expansion of the fractional kinetic operator, in \cite{Barci:1995ad,Barci:1996br,Barci:1996ny,Barci:1998wp}.

The free scalar field discussed in \cite{Frassino:2019yip}, derived from the fractional scalar field through the extension problem, falls within the category of non-local field theories featuring fractional operators. These theories have attracted considerable interest, particularly in addressing pivotal issues within quantum field theories, quantum gravity in the UV regime (refer to \cite{Calcagni:2021ljs} for a recent review), and their relevance in string theory \cite{Erbin:2021hkf,Nortier:2021six}.

The introduction of non-locality through fractional operators, as extensively discussed in \cite{Calcagni:2022shb}, facilitates the exploration of local interaction terms. These terms, stemming from non-local scalar field theories with fractional Laplacians, are often more straightforward and manageable in comparison to interactions arising from delocalization, a concept extensively investigated in \cite{Tomboulis:2015gfa}. Furthermore, amplitudes stemming from non-local interaction vertices have been observed in string field theory \cite{Pius:2016jsl,Pius:2018crk,Chin:2018puw}. The non-local scalar field theories with fractional Laplacians, akin to the focus of this study, have recently found applications in constructing MOND models \cite{Giusti:2020rul,Giusti:2020kcv} and calculating entanglement entropy in long-range interacting systems \cite{Roy:2021akq}. Investigations into massless and massive non-local scalar field theories within the context of non-local conformal field theories and the entanglement entropy problem are presented in \cite{Basa:2019ywr,Basa:2020cyn}. Moreover, the extension problem has been leveraged to scrutinize anomalous dimensions for boundary conserved currents in holography, encompassing general $p$-forms and Maxwell equations \cite{LaNave:2017nex}, and to explore fractional electromagnetism \cite{LaNave:2019mwv,Heydeman:2022yni}.

In this paper, we explore the classical aspects of the lowest-dimensional realization of the extension problem \cite{Caffarelli:2007}, specifically focusing on the mapping between a one-dimensional non-local theory on the boundary and a two-dimensional local field theory in the bulk, with a particular emphasis on applications to classical mechanics. To establish a physically meaningful representation of this mapping, we introduce a model of the classical fractional particle. This model considers $n$ independent scalar fields representing the coordinates of the particle in Euclidean $\mathbb{R}^n$ space, with the one-dimensional variable interpreted as the time-variable. The significance of the fractional particle model lies in its generalization of the classical particle to the fractional realm, a domain where the fractional quantum one-particle Schr\"{o}dinger equation has already been under investigation for some time \cite{Laskin:1999tf}. Additionally, it extends the understanding of the non-local particle, with a kinetic term defined by the standard derivative interpreted in terms of strings \cite{Cheng:2008qz}. An insightful observation, presented in \cite{Calcagni:2013eua}, asserts that the Nambu-Goto action can be approximated by a non-local particle action when examining strings from a large distance. This observation underscores the fundamental association of non-locality with extended objects.

While non-locality is a fundamental aspect of quantum mechanics, justifying the examination of fractional quantum particles, it represents a conceptual challenge to make sense of the corresponding classical fractional particle model. This paper aims to contribute to this problem, by introducing and discussing the fractional particle, along with some of its physical properties. Specifically, we explore the relationship between momentum and kinetic energy, showcasing that the latter can be expressed in terms of both instantaneous momentum, viewed as the product of instantaneous velocity and mass, and canonical momentum. This formulation shows some light on the complex challenges inherent in the construction of the Hamiltonian formalism for field models governed by fractional Laplacians. Further discussions encompass the symmetries of the free fractional particle model, the equations of motion, and the application of the general method of Green's function to solve these equations. We present three illustrative examples of fractional particles: the free particle, the fractional harmonic oscillator and the charged fractional particle in an electromagnetic field. Applying the results of the extension problem to the fractional particle, we establish a mapping to a two-dimensional classical sigma model on the positive half-plane. The extension problem and the particle equation of motion determine the boundary conditions imposed on the local sigma model in the bulk. We solve the classical equations of motion with fractional particle boundary conditions for the sigma model fields. Subsequently, we quantize the sigma model using the canonical quantization method, deriving the canonical commutation relations of the mode operators, the canonical Hamiltonian, and calculating the vacuum energy for any value of the fractional parameter.

Although our construction of fractional particles is grounded in the field theory perspective, it is crucial to highlight the broader applicability of the fractional particles across various physics problems. Fractional calculus, gaining attention in recent years, offers a means to model complex physical phenomena beyond the reach of standard calculus. Since there are different definitions of the fractional operators, different types of fractional particle models can be constructed. Various perspectives provide motivation for employing the fractional particle model with the fractional Laplacian in the kinetic term. Mathematically, the inclusion of the fractional Laplacian in the kinetic term facilitates the implementation of the extension problem by Caffarelli and Silvestre, enabling a straightforward mapping of the particle to a non-fractional two-dimensional field theory. From an applied standpoint, fractional particle models hold a great potential across various fields. Specifically, the fractional Laplacian allows the generalization of diffusion and random walks, making it pertinent to systems with long-range interactions and correlations. As an instance of applications, the classical fractional particle model finds relevance in the study of systems characterized by fractional Brownian motion, where statistical properties exhibit long-range correlations \cite{Bock:2020}. Additionally, the model can be applied to materials with viscoelastic properties, such as certain polymers and soft tissues, whose behavior eludes full capture by classical models. The dynamics of particles within these materials, accounting for non-local interactions and memory effects inherent in the fractional Laplacian formalism, can be elucidated through classical fractional particles \cite{Failla:2020}. Furthermore, the model provides a fine-grained description of heat conduction in materials with anomalous thermal behavior, such as fractal structures or materials with long-range correlations \cite{Zecova:2015} and turbulent flows \cite{Suzuki:2023}. Finally, the classical fractional particle offers insight into fractional diffusion in porous media. In such cases, the model can effectively study how particles diffuse through porous structures with long-range interactions and fractional derivatives \cite{Vasquez:2012}.

The structure of this paper is as follows: In Section 2, we present the fractional classical particle model featuring the fractional Laplacian. The construction method is straightforward, aligning with the principles of fractional operator equations, wherein the standard Laplacian is replaced by the fractional Laplacian in the kinetic term. This incorporation of the fractional Laplacian introduces non-locality into the theory at the level of free particles, allowing us to keep the interactions local. This section encompasses discussions on equations of motion, linear momentum, kinetic energy, symmetries of the free particle, and the application of the Green's function method in the general case. In Section 3, we analyse three illustrative examples: the free fractional particle, the fractional harmonic oscillator, and the electrically charged fractional particle in an electromagnetic field. Section 4 employs the extension problem to construct the corresponding classical local sigma model. We derive the equations of motion in the bulk and achieve exact solutions, building upon the results established in \cite{Frassino:2019yip} for the case of free scalar fields. Subsequently, we undertake the quantization of the sigma model using the canonical quantization method. This involves determining the canonical Hamiltonian and presenting an explicit formula for the vacuum energy. In the last section, we discuss the results obtained and outline potential avenues for further research inspired by the insights gained in this work.
In the Appendix A, we have compiled some definitions of the fractional Laplacian, and in the Appendix B, we have given the formulas of the vacuum energy at the boundary and the plots of vacuum energy for some values of the fractionality parameter.

%-----------------------------------------------------------------
\section{Fractional Classical Particle} 
\label{sec:nlsf}

In this section, we introduce the fractional classical particle in $\mathbb{R}^n$ as a natural extension of the Newtonian particle. This novel model is derived by replacing the one-dimensional Laplacian, corresponding to the second derivative with respect to the time variable, with the fractional Laplacian. We examine the key aspects of this model, including the definition of momentum and kinetic energy, the global symmetries, and the application of the Green's function method to solve the equations of motion. Additionally, we investigate the behavior of the fractional particle within a generalized force field.

%-----------------------------------------------------------------
%-----------------------------------------------------------------
\subsection{Equations of Motion of Fractional Particle}
\label{sec:nlsf-eom}

We start by recalling the action of the classical particle that moves in a $n$-dimensional Euclidean space under the influence of a potential $V[x]$, given by 
\begin{equation}
S[x] = \int_{-\infty}^{+\infty} d\tau \left[ \frac{m}{2} \frac{d x^a (\tau)}{d \tau} \frac{d x^b (\tau )}{d \tau} \eta_{ab} - V[x] \right]
\, .
\label{fsm-fnp-act-newt}
\end{equation} 
Here, $m$ is the particle mass, $x^a : \mathbb{R} \rightarrow \mathbf{R}^n$ are smooth maps on the time parameter $\tau \in \mathbb{R}$ to the target-space $\mathbf{R}^n$, the Latin indices $a, b, \ldots = 0, 1, \ldots , n-1$ indicate the components of the target-space objects, and $V[x]$ is a potential functional that is smooth in all variables. Also, we denote by $\eta_{ab} = \delta_{ab}$ the components of the Euclidean metric tensor and we use the covariant notation for convenience. 
The kinetic term from $S[x]$ can be rewritten as
\begin{equation}
S_0 [x] = - \frac{m}{2} \int_{-\infty}^{+\infty} d\tau \, x^a (\tau) 
\frac{d^2 x^b (\tau )}{d \tau^2} \eta_{ab}
\, ,
\label{fsm-fnp-act-newt-0}
\end{equation} 
if $x^a$'s satisfy the following boundary conditions
\begin{equation}
x_a (\tau) \frac{d x^a (\tau )}{d \tau} \to 0 \quad \text{as} \quad \tau \to \pm \infty
\, .
\label{fsm-fnp-act-newt-bc}
\end{equation} 

The particle model discussed above can be extended to a non-local particle model. This can be achieved by introducing non-locality in one of two ways: either by substituting the second derivative in the kinetic term with a fractional second order derivative (specifically, the one-dimensional fractional Laplacian), or by delocalizing the interactions from $V[x]$ (refer to \cite{Tomboulis:2015gfa} for more details), or potentially both. This paper will focus on exploring the first method, which results in the free fractional particle action
\begin{equation}
S^{(\alpha)}_0 [x] = - \frac{m_{\alpha}}{2} \int_{-\infty}^{+\infty} d\tau \, x^a (\tau) \left( - \Delta_{\tau} \right)^{\frac{\alpha}{2}} x^b (\tau ) \eta_{ab} 
\, .
\label{fsm-fnp-act-newt-01}
\end{equation} 
The parameter $m_{\alpha}$ formally plays the similar role in $S^{(\alpha)}_0 [x]$ as the particle mass $m$ in the action $S_0 [x]$ with which $m_{\alpha}$ must coincide at $\alpha = 2$. However, in natural units $c = \hbar = 1$, the dimension of $\langle m_{\alpha} \rangle = E^{3-\alpha}$ for any arbitrary value of $\alpha$. That shows that the simplest choice for $m_\alpha$ is $m_\alpha = m^{3-\alpha}$. Since we want to describe a fractional classical particle, we interpret the one-dimensional variable $\tau$ from $(-\Delta_{\tau})^{\frac{\alpha}{2}}$ as a time-variable. With that, our model describes the evolution of the fractional particle in the $n$-dimensional Euclidean space. Also, it describes $n$ one-dimensional scalar fields. The one dimensional fractional Laplacian is equivalent to the Riesz derivative on $\mathbb{R}$ or, in general, on $\mathbb{R}^n$. In what follows, we consider the coordinates $x^a \in \mathscr{X}$ where $\mathscr{X}$ is any of the spaces $\mathscr{L}^p$, $\mathscr{C}_0$, $\mathscr{C}_{b u}$ or $\mathscr{S}$ on which the definitions of the fractional Laplacian given in the Appendix A are equivalent. As in the case of the standard particle model, we can incorporate local interaction terms into the kinetic action $S^{(\alpha)}_0 [x]$ to depict the dynamics of the fractional particle in an external field
\begin{equation}
S^{(\alpha)} [x] = - \int_{-\infty}^{+\infty} d\tau \, 
\left[ \frac{m_{\alpha}}{2} x^a (\tau) \left( - \Delta_{\tau} \right)^{\frac{\alpha}{2}} x^b (\tau ) \eta_{ab} 
+
V[x]
\right]
\, .
\label{fsm-fnl-act-frac-newt-1}
\end{equation}
If $V[x]$ is an arbitrary polynomial, $S^{(\alpha)} [x]$ characterizes a non-linear non-local fractional particle model. 

The equation of motion of the fractional particle can be obtained by applying the variational principle to the action $S^{(\alpha)} [x]$. To this end, consider the arbitrary infinitesimal variation of the coordinates
\begin{equation}
x^a (\tau ) \rightarrow x^{\prime a} (\tau ) = x^{a} (\tau ) + \delta x^a (\tau )
\, .
\label{fsm-fnl-eom-1}
\end{equation}  
The variation of $S^{(\alpha)} [x]$ under the transformations (\ref{fsm-fnl-eom-1}) is given by
\begin{equation}
\delta S^{(\alpha)} [x] = - \int_{-\infty}^{+\infty} d\tau \, 
\left[ 
\frac{m_{\alpha}}{2} 
\left( \delta x^a (\tau) \left( - \Delta_{\tau} \right)^{\frac{\alpha}{2}} x_a (\tau )
+
x^a (\tau) \left( - \Delta_{\tau} \right)^{\frac{\alpha}{2}} \delta x_a (\tau )
\right)
+
\frac{\partial V[x]}{\partial x^a (\tau) } \delta x^a (\tau) ]
\right]
\, .
\label{fsm-fnl-eom-2}
\end{equation}
Next, by using the following inversion property of the fractional Laplacian distribution
\begin{equation}
\int_{-\infty}^{+\infty} f(\tau ) \left(- \Delta_{\tau} \right)^{\frac{\alpha}{2}} g(\tau ) d \tau =
\int_{-\infty}^{+\infty} g(\tau ) \left(- \Delta_{\tau} \right)^{\frac{\alpha}{2}} f(\tau ) d \tau
\, ,
\label{fsm-fnp-L-frac-tr-3}
\end{equation}
for any two functions $f$ and $g$ that satisfy the condition (\ref{fsm-fnp-L-frac-def-0}), we obtain the equation of motion
\begin{equation}
m_{\alpha} \left(-\Delta_\tau \right)^{\frac{\alpha}{2}} x^{a} (\tau) 
+ \frac{\partial V[x]}{\partial x_a (\tau) } = 0
\, .
\label{fsm-fnp-L-frac-eom-3}
\end{equation}
Equation (\ref{fsm-fnp-L-frac-eom-3}) represents a specific instance of the fractional Poisson equation, which is occasionally referred to as the fractional Laplace equation, with $\partial^a V[x]$ serving as the source. Generally, the Poisson equation can accommodate more general sources $f^{a} (\tau )$, which may not necessarily depend on $x^a$. These sources are incorporated into the model based on phenomenological reasoning rather than the action principle. Interpreted as generalized forces, these sources give rise to a non-homogeneous, non-local equation of motion
\begin{equation}
m_{\alpha} \left(-\Delta_\tau \right)^{\frac{\alpha}{2}} x^{a} (\tau) 
+ \frac{\partial V[x]}{\partial x_a (\tau) } = f^{a} (\tau )
\, .
\label{fsm-fnp-L-frac-eom-4}
\end{equation}

Although the fractional particle model presented above appears to be a straightforward extension of the standard particle model, it's important to acknowledge that the properties of $(-\Delta_{\tau})^{\frac{\alpha}{2}}$ render this generalization non-trivial, both mathematically and physically. To further elucidate this point, we will discuss the linear momentum and kinetic energy of fractional particles.

%-----------------------------------------------------------------
%-----------------------------------------------------------------
\subsection{Linear Momentum and Energy}
\label{sec:nlsf-lme}

Generally, the task of defining the linear or canonical momentum for fractional systems is complex, primarily because the fractional Laplacian lacks a straightforward interpretation in the context of tangent spaces. Furthermore, given that the kinetic energy exhibits non-locality in time, it is anticipated that the momentum will also be non-local. To understand how these concepts are embodied within the previously discussed fractional particle model, we need to first examine the relationship between the operator $\left(-\Delta_\tau \right)^{\frac{\alpha}{2}}$ and the conventional first derivative $d/d \tau$. For this purpose, we employ the equivalent formulation of the fractional Laplacian in terms of the regularized principal value, as per \cite{Kwasnicki:2017}, which leads us to the following expression
\begin{align}
\left(-\Delta_\tau \right)^{\frac{\alpha}{2}} x^{a} (\tau ) & = C_{1,\alpha}  
\int_{0}^{+\infty} d \zeta \frac{x^{a}(\tau + \zeta)+ x^{a}(\tau - \zeta) - 2 x^{a}(\tau )}{\zeta^{1+\alpha}}
\nonumber
\\
& = - \frac{C_{1,\alpha}}{\alpha} \frac{d}{d \tau}
\left[ 
\int_{0}^{+\infty} d \zeta \frac{x^{a}(\tau + \zeta)- x^{a}(\tau - \zeta)}{\zeta^{\alpha}} 
\right]
\, ,
\label{flm-L-def-reg}
\end{align}
for all $0 \leq \alpha < 2$. The above relation can be used to define the first-order fractional derivative as
\begin{equation}
\left( - \Delta^{(1)}_\tau \right)^{\frac{\alpha}{2}} x^{a} (\tau )= 
c_{1,\alpha} 
\left[ 
\int_{0}^{+\infty} d \zeta \frac{x^{a}(\tau + \zeta)- x^{a}(\tau - \zeta)}{\zeta^{\alpha}} 
\right]
\, ,
\label{flm-L-def-reg-1}
\end{equation}
where $c_{1,\alpha}= - C_{1,\alpha}/ \alpha$. The equation (\ref{flm-L-def-reg}) gives the relationship between the fractional Laplacian, the first derivative, and the first-order fractional derivative, which can be formally written as
\begin{equation}
\left(-\Delta_\tau \right)^{\frac{\alpha}{2}} = \frac{d}{d \tau} \left[ \left( - \Delta^{(1)}_\tau \right)^{\frac{\alpha}{2}} \right]
\, .
\label{flm-L-def-reg-2} 
\end{equation}
The above relations can be used to recast the free particle action from the equation (\ref{fsm-fnp-act-newt-01}) into the following form
\begin{equation}
S^{(\alpha)}_0 [x] = - \frac{m_{\alpha}}{2} \int_{-\infty}^{+\infty} d\tau \, x^a (\tau) \left( - \Delta_{\tau} \right)^{\frac{\alpha}{2}} x^b (\tau ) \eta_{ab} =
\frac{m_{\alpha}}{2} 
\int_{-\infty}^{+\infty} d\tau \, \frac{d x^a (\tau)}{d \tau } \left( - \Delta^{(1)}_{\tau} \right)^{\frac{\alpha}{2}} x^b (\tau ) \eta_{ab} 
\, .
\label{fsm-fnp-act-newt-01-}
\end{equation}
From $S^{(\alpha)}_0 [x]$, we can derive the canonical momentum 
\begin{equation}
\pi^{a}_{\alpha} (\tau ) = \frac{\partial L}{\partial (\partial_{\tau} x_{a} (\tau ))} = 
\frac{m_{\alpha}}{2} 
\left( - \Delta^{(1)}_{\tau} \right)^{\frac{\alpha}{2}} x^a (\tau ) 
\, . 
\label{fsm-fnp-can-mom}    
\end{equation}
By introducing the classical instantaneous linear momentum $\mathbf{p} (\tau ) = m \mathbf{v} (\tau ) = m \dot{\mathbf{x}} (\tau )$, the kinetic energy at any instant $\tau$ can be written as
\begin{equation}
T_{\alpha} (\tau ) = \frac{p_{a} (\tau) \pi^{a}_{\alpha} (\tau )}{2 m}
\, .
\label{fsm-fnp-kin-energ}
\end{equation}
The expression (\ref{fsm-fnp-kin-energ}) extends the notion of kinetic energy to the fractional particle model. Unlike the classical scenario, the relationship between the tangent vector to the trajectory at a given parameter $\tau$ and the energy is no longer dictated by a scalar product within either the tangent space or the phase space. This deviation arises due to the involvement of the operator $\left( - \Delta^{(1)}_{\tau} \right)^{\frac{\alpha}{2}}$, which does not reside in either of these conventional spaces. Remarkably, in the limit approaching classical dynamics, the traditional formula for kinetic energy is recovered
\begin{equation}
\lim_{\alpha \rightarrow 2} T_{\alpha} (\tau) = 
\frac{m \dot{x}_a (\tau ) \dot{x}^a (\tau )}{2} = \frac{p_{a}(\tau )p^{a}(\tau )}{2 m}
\, .
\label{fsm-fnp-kin-energ-1}
\end{equation}
In the presence of a potential or interaction term, the total particle energy is the sum between its kinetic and potential energies, as in classical mechanics. The conservation principles of energy and linear momentum in classical mechanics stem from the model's invariance under time and space translations, as elucidated by Noether's theorem. Consequently, the preservation of these fundamental properties warrants careful consideration when extending the analysis to the realm of fractional particles.

We note that the structure of $T_{\alpha} (\tau)$ is determined by the necessity for the kinetic term to be represented in terms of the fractional Laplacian. Given that the kinetic energy no longer exhibits quadratic dependence on momentum, a natural inclination arises to construct the Lagrangian employing a first-order fractional derivative, resembling $ \sim \left( - \Delta^{(1)}_{\tau} \right)^{\frac{\alpha}{2}} x_a (\tau ) \left( - \Delta^{(1)}_{\tau} \right)^{\frac{\alpha}{2}} x^a (\tau )$. This alternative formulation leads to a distinctive fractional particle model, where the conventional $d/d \tau$ in the Newtonian particle action is substituted by $\left( - \Delta^{(1)}_{\tau} \right)^{\frac{\alpha}{2}}$. However, despite the symmetrical expression in $x^a (\tau )$'s, establishing a connection between the kinetic energy and canonical momentum becomes unfeasible, as the latter is no longer well defined in this model.  Furthermore, the interpretation of $\left( - \Delta^{(1)}_{\tau} \right)^{\frac{\alpha}{2}} x^a (\tau )$ remains ambiguous, complicating the understanding of this model. Despite these limitations, the fractional particle model resulting from this substitution warrants investigation. Nonetheless, due to its deviation from the extension problem by Caffarelli and Silvestre, it will not be further elaborated upon in this paper.

%-----------------------------------------------------------------
%-----------------------------------------------------------------
\subsection{Symmetries of Fractional Particle Model}
\label{sec:nlsf-sym}

Let us examine the symmetries of the free fractional particle action given by the equation (\ref{fsm-fnp-act-newt-01}). By construction, the action $S^{(\alpha)}_0 [x]$ is invariant under rotations within the target-space, denoted as
\begin{equation}
x^a (\tau ) \to x^{\prime a} (\tau ) = {\Lambda^a}_b \, x^b (\tau ) 
\, , 
\label{fsm-fnp-L-frac-rot}
\end{equation}
where $\bm{\Lambda} \in  SO(n)$. The translations within the target space are characterized by transformations of the form
\begin{equation}
x^a (\tau ) \to x^{\prime a} (\tau ) = x^a (\tau ) + {\epsilon}^a
\, , 
\label{fsm-fnp-L-frac-tran}
\end{equation}
where $\bm{\epsilon} \in \mathbb{R}^n$ represents a constant vector field. The variation of $S^{(\alpha)}_0 [x]$ under such translations, denoted by 
(\ref{fsm-fnp-L-frac-tran}), is expressed as
\begin{align}
\delta_{\epsilon} S^{(\alpha)}_0 [x] 
& = S^{(\alpha)}_0 [x^{\prime}] - S^{(\alpha)}_0 [x] 
=
- \frac{m_{\alpha}}{2} \epsilon_{a}\int_{-\infty}^{+\infty} d\tau  \left( - \Delta_{\tau} \right)^{\frac{\alpha}{2}} x^a (\tau)
\nonumber
\\
& = - \frac{m_{\alpha}}{2} \epsilon_{a} 
\left. \left( - \Delta^{(1)}_{\tau} \right)^{\frac{\alpha}{2}} x^a (\tau) \right\vert_{-\infty}^{+\infty} = 0
\, ,
\label{fsm-fnp-L-frac-tran-1}
\end{align}
where, in the second line above, we used equation (\ref{flm-L-def-reg-1}) and the fact that
\begin{equation}
\lim_{| \tau | \to \infty} \left( - \Delta^{(1)}_{\tau} \right)^{\frac{\alpha}{2}} x^a (\tau) = 0
\, ,
\label{fsm-fnp-L-frac-tran-2}
\end{equation}
for $0 < \alpha < 2$. It's important to note that the last equality in (\ref{fsm-fnp-L-frac-tran-1}) holds as a consequence of the Laplacian operator's definition. In the case where $\alpha = 0$, we obtain
\begin{equation}
\int_{-\infty}^{+\infty} d\tau \left( - \Delta^{(1)}_{\tau} \right)^{0} \, x^a (\tau)
= \int_{-\infty}^{+\infty} d\tau  \, x^a (\tau)
\, ,
\label{fsm-fnp-L-frac-tran-3}
\end{equation}
which is expected to vanish in accordance with the properties of the coordinates $x^{a}(\tau ) \in \mathscr{X}$. 

Let us now examine the implications of world-line translations. We consider infinitesimal transformations of the time variable given by
\begin{equation}
\tau \to \tau^{\prime} = \tau + \delta \tau = \tau + \epsilon
\, ,
\label{fsm-fnp-L-frac-tr}
\end{equation}
where $\epsilon$ represents an infinitesimal real constant parameter. The variation of $S^{(\alpha)}_0 [x]$ under these transformations (\ref{fsm-fnp-L-frac-tr}) is conventionally defined as
\begin{equation}
\delta S^{(\alpha)}_0 [x(\tau )] =  S^{(\alpha)}_0 [x'(\tau + \epsilon)] - S^{(\alpha)}_0 [x(\tau )]
\, .
\label{fsm-fnp-L-frac-tr-1}
\end{equation}
As the coordinates are scalar functions $x^{\prime a}(\tau^{\prime}) = x^{a}(\tau)$, it is imperative to establish the invariance of the fractional Laplacian under world-line translations. Utilizing the integral representation of the fractional Lagrangian, as defined by equation (\ref{fsm-fnp-L-frac-def-1}), we can compute the variation of the fractional Laplacian resulting from these world-line translations as follows
\begin{equation}
\left( - \Delta_{\tau^{\prime}} \right)^{\frac{\alpha}{2}} x^{\prime b} (\tau^{\prime} ) =
C_{1,\alpha} \int_{- \infty}^{+\infty} d \zeta^{\prime}
\frac{ x^{\prime b} (\tau^{\prime}) - x^{\prime b} (\zeta^{\prime})}{\vert \tau^{\prime} - \zeta^{\prime} \vert^{1+\alpha}} 
= \left( - \Delta_{\tau} \right)^{\frac{\alpha}{2}} x^b (\tau )
\, . 
\label{fsm-fnp-L-frac-tr-2}
\end{equation}
Substituting (\ref{fsm-fnp-L-frac-tr-2}) into (\ref{fsm-fnp-L-frac-tr-1}), we arrive at
\begin{equation}
\delta S^{(\alpha)}_0 [x(\tau )] = 0
\, .
\label{fsm-fnp-L-frac-tr-4}
\end{equation}
This discussion shows that the fractional particle exhibits the expected symmetries, which converge to the known symmetries of the classical particle as $\alpha \to 2$. 

Polynomial interaction potentials, as mentioned previously, can be incorporated into the framework. To uphold rotational symmetry, these potentials should remain functions of $V[x^2] = V[x^a x_a]$ in the conventional manner. On the other hand, potentials explicitly dependent on $\tau$ do not maintain reparametrization invariance.

%-----------------------------------------------------------------
%-----------------------------------------------------------------
\subsection{Fractional Green's Function}
\label{sec:nlsf-green}

The dynamics of fractional particles can be better understood through an examination of their equations of motion. For a large class of systems where the particle is subjected to generalized forces, solutions to the equations of motion are effectively obtained through the application of the method of time-dependent Green's functions. It's important to note, however, that the Green's functions of the fractional Laplacian depend on the function space considered, as well as on the imposed boundary conditions. In our analysis, we assume that the functions involved belong to any of the spaces denoted by $\mathscr{X}$. As established in \cite{Kwasnicki:2017}, all three definitions of the fractional Laplacian provided in the Appendix A are equivalent on $\mathscr{X}$. 

Under this hypothesis, let us consider equation (\ref{fsm-fnp-L-frac-eom-4}) with generalized sources $f^{a}(\tau ) = s^{a} (\tau )$. Here, we consider a constant potential $\partial V[x]/ \partial x^a = 0$, although the results can be readily extended to non-constant potentials. The fractional Green's function, denoted as $G_{\alpha} (\tau , \tau^{\prime})$, satisfies the associated fractional Green's equation
\begin{equation}
\left(-\Delta_\tau \right)^{\frac{\alpha}{2}} G_{\alpha} (\tau , \tau^{\prime})
= \delta (\tau - \tau^{\prime})
\, , 
\label{fsm-fnp-L-frac-Green-1}
\end{equation}
where $\tau$ and $\tau^{\prime}$ belong to $\mathbb{R}$, and we assume that  $G_{\alpha} (\tau , \tau^{\prime}) = G_{\alpha} (\tau - \tau^{\prime})$ are invariant under translations. 

The equation (\ref{fsm-fnp-L-frac-Green-1}) is an equality in the sense of distributions acting on test functions $\varphi$, as specified in equation (\ref{fsm-fnp-L-frac-def-2}). Since the Fourier and distributional definitions are equivalent, $\left(-\Delta_\tau \right)^{\frac{\alpha}{2}} G_{\alpha} (\tau - \tau^{\prime})$ can be defined as outlined in equation (\ref{fsm-fnp-L-frac-def}) \cite{Kwasnicki:2017}. Exploiting this equivalence, for any fixed $\tau^{\prime}$ and any test function $\varphi$, we obtain
\begin{align}
\int_{-\infty}^{+\infty} d \tau \, \varphi (\tau) \left(-\Delta_\tau \right)^{\frac{\alpha}{2}} G_{\alpha} (\tau - \tau^{\prime}) 
& =
\int_{-\infty}^{+\infty} d \tau \, \varphi (\tau) 
\mathcal{F}^{-1} \left\{ |\omega |^{\alpha} \mathcal{F}\{G_{\alpha} \} (\omega ) \right\} (\tau - \tau^{\prime})
\nonumber
\\
& =
\int_{-\infty}^{+\infty} d \tau \, \varphi (\tau) 
\delta (\tau - \tau^{\prime}) 
\, .
\label{fsm-fnp-L-frac-Green-2}
\end{align}
Utilizing the elementary properties of the Fourier transforms of distributions, where $\delta = \mathcal{F}^{-1}\{ \mathds{1} \}$, we arrive at
\begin{equation}
\tilde{G}_{\alpha} (\omega) := \mathcal{F}\{G_{\alpha}(\tau - \tau^{\prime}) \} (\omega ) = \frac{1}{|\omega |^{\alpha}}
\, .
\label{fsm-fnp-L-frac-Green-3}
\end{equation}

By applying the inverse Fourier transform (\ref{fsm-fnp-L-frac-Fou-inv}) to the equation (\ref{fsm-fnp-L-frac-Green-2}), we obtain the following Green's function
\begin{equation}
{G}_\alpha(\tau - \tau^{\prime})=
\int_{-\infty}^{\infty} \frac{d \omega}{2 \pi} 
\frac{1}{|\omega|^{\alpha}} e^{i \omega (\tau - \tau^{\prime})}
= \frac{1}{\pi} \sin \left(\frac{\pi \alpha}{2} \right) 
\Gamma(1-\alpha)|\tau- \tau^{\prime}|^{\alpha-1} 
\, .
\label{fsm-fnp-L-frac-Green-4}
\end{equation}
Since the fractional Laplacian is a linear operator, the general solutions to the equation of motion with arbitrary sources can be constructed using the Green's function 
\begin{equation}
x^{a}_{\alpha} (\tau ) = x^{a}_{\alpha , 0} (\tau ) +
\frac{1}{\pi m_{\alpha}} \sin \left(\frac{\pi \alpha}{2} \right) 
\Gamma(1-\alpha)
\int_{-\infty}^{+\infty} d \tau^{\prime}  
\frac{1}{|\tau - \tau^{\prime}|^{1-\alpha}} s^{a} (\tau^{\prime})
\, ,
\label{fsm-fnp-L-frac-Green-5}
\end{equation}
where $x^{a}_{\alpha , 0} (\tau )$ is a solution of the homogeneous fractional Poisson equation. The instantaneous particle velocity $\mathbf{v}(\tau )$ is the tangent vector to the trajectory at $\tau$ which can be calculated from (\ref{fsm-fnp-L-frac-Green-5}), and it is given by
\begin{equation}
v^{a}_{\alpha} (\tau ) = v^{a}_{\alpha , 0} (\tau ) +
\frac{\alpha-1}{\pi m_{\alpha}} 
\sin \left(\frac{\pi \alpha}{2} \right) 
\Gamma(1-\alpha)
\int_{-\infty}^{+\infty} d \tau^{\prime}  
\frac{\tau - \tau^{\prime}}{|\tau - \tau^{\prime}|^{3-\alpha}}
s^{a} (\tau^{\prime})
\, ,
\label{fsm-fnp-L-frac-Green-6}
\end{equation}
where $v^{a} (\tau ) = d x^{a} (\tau) / d \tau$. This vector determines $\mathbf{p} (\tau ) = m \mathbf{v}(\tau )$ as in the classical mechanics. Its connection with the fractional particle's energy has been discussed in subsection \ref{sec:nlsf-lme}.

Some comments are in order here. Since we have interpreted $\tau$ as being the time-variable, the solution (\ref{fsm-fnp-L-frac-Green-6}) describes the instantaneous position of the fractional particle moving in the generalized time-dependent force field $\mathbf{s} (\tau )$. Nevertheless, at any given instant of time, the particle position depends on the value of $\mathbf{s} (\tau )$ at all other instants of time, which is a consequence of the non-local (in time) character of the equation of motion (\ref{fsm-fnp-L-frac-eom-4}). The last term of the equation (\ref{fsm-fnp-L-frac-Green-5}) resembles the non-local functions in time that are employed in the fractional calculus in the description of systems with memory (see, e. g. \cite{Singh:2022}). However, there is a major distinction between the two cases because the equation (\ref{fsm-fnp-L-frac-Green-5}) calculates contributions made by the generalized force field from both the past and the future. 

It is important to observe that, as in the case of standard linear operators, the Green's functions of $\left(-\Delta_\tau \right)^{\frac{\alpha}{2}}$ depend on the boundary conditions of the problem. To illustrate this point, let us consider finite time intervals $\tau \in [\tau_1 , \tau_2 ]$ and the boundary conditions 
\begin{equation}
G_{\alpha}(\tau_1, \tau^{\prime}) = G_{\alpha}(\tau_2, \tau^{\prime}) = 0
\, ,
\label{fsm-fnp-L-Green-7}
\end{equation}
where $\tau^{\prime}$ is a singular point. The equation (\ref{fsm-fnp-L-Green-7}) suggests that the Green's function can be expanded in terms of sinusoidal functions as follows
\begin{equation}
G_{\alpha} \left(\tau, \tau^{\prime} \right)
= \sum_{n=1}^{\infty} C_n \left(\tau^{\prime} \right) 
\sin \left[ \frac{n \pi (\tau - \tau_1) }{\tau_2 - \tau_1} \right]
\, .
\label{fsm-fnp-L-frac-Green-s-1}
\end{equation} 
After substituting the right hand side of the equation (\ref{fsm-fnp-L-frac-Green-s-1}) into the equation (\ref{fsm-fnp-L-frac-Green-1}), and after some algebraic manipulations, we obtain
\begin{equation}
G_\alpha \left( \tau, \tau^{\prime} \right)
= \frac{2}{\tau_1 - \tau_2} \sum_{n=1}^{\infty}
\left(\frac{\tau_2 - \tau_1}{n \pi}\right)^\alpha 
\sin \left[ \frac{n \pi (\tau - \tau_1)}{\tau_2 - \tau_1}
\right] 
\sin \left[ \frac{n \pi (\tau^{\prime}- \tau_1 )}{\tau_2 - \tau_1} \right]
\, .
\label{fsm-fnp-L-frac-Green-s-2}
\end{equation}
In determining the above Green's function which is symmetric, the boundary conditions played a crucial role. 

As a final comment, the Green's function is not the only method to calculate the solutions to the equations of motion. If the particle movement is limited to a finite interval of time, different methods can be employed to study the fractional Poisson equation. For more details, see \cite{Chen:2020}.

%-----------------------------------------------------------------
\section{Examples of Fractional Particle Models}
\label{sec:exfp}

This section explores fractional particle models that extend classical models, as discussed previously. The challenge of obtaining general solutions for fractional Poisson equations, whether on $\mathbb{R}^d$ or within an open ball $B \in \mathbb{R}^d$, where $d$ represents the number of variables, is currently a subject of active research in mathematics. For the fractional particle, with $d=1$ and $B = I \in \mathbb{R}$ representing an open real interval, the boundary conditions, serving as initial and final conditions, define the pseudo-differential problem defining the particle dynamics.

%-----------------------------------------------------------------
%-----------------------------------------------------------------
\subsection{Free Fractional Particle}
\label{sec:exfp-free}

The equation governing the motion of a fractional particle moving in the absence of external influence is described by equation (\ref{fsm-fnp-L-frac-eom-3}) with $V[x] = 0$, yielding the fractional Laplace equation
\begin{equation}
\left(-\Delta_{\tau} \right)^{\frac{\alpha}{2}} x^{a} (\tau) = 0
\, .
\label{fsm-fnp-frac-Lapl-1}
\end{equation}
Given that $\left(-\Delta_{\tau} \right)^{\frac{\alpha}{2}}$ is a non-local operator, the fractional particle position at an instant of time is determined by its positions at all instants of time. To illustrate the distinction between the dynamics of free fractional and standard particles, consider the scenario where the particle is initially stationary at the origin of $\mathbb{R}^{n}$ before $\tau_1$ and after $\tau_2 > \tau_1$. From a mathematical perspective, this scenario defines a Dirichlet problem for the fractional Laplace equation
\begin{align}
\left(-\Delta_{\tau} \right)^{\frac{\alpha}{2}} x^{a} (\tau) & = 0  
\quad \text{for} \quad \tau \in (\tau_1,\tau_2) \subseteq \mathbf{R}
\, ,
\label{fsm-fnp-frac-Lapl-2-a}
\\
x^a (\tau )  & = 0
\quad \text{for} \quad \tau \in \mathbf{R} \backslash [\tau_1,\tau_2]
\, .
\label{fsm-fnp-frac-Lapl-2-b}  
\end{align}
The solution to the Dirichlet problem (\ref{fsm-fnp-frac-Lapl-2-a}) and (\ref{fsm-fnp-frac-Lapl-2-b}) can be expressed in terms of $s$-harmonic functions, which are defined as follows \cite{Landkof:1972}: A function 
$h_s : \mathbb{R} \rightarrow \mathbb{R}_+$, continuous in $I=(\tau_1,\tau_2) \subseteq \mathbb{R}$, is $s$-harmonic on $I$ if it satisfies the equation
\begin{equation}
h_s(x) = 
\int_{\mathbb{R} \backslash I_{\xi}(\tau)}
\frac{C(1, s) \xi^{2 s}}{ \vert \zeta - \tau \vert 
\left(
	|\zeta - \tau|^2 - \xi^2
\right)^s} 
h_s(\zeta ) d \zeta 
\, .
\label{fsm-fnp-frac-s-harm}
\end{equation}
Here, the punctured zero-dimensional disk $I_{\xi} (\tau)$ is fully contained in the open interval $I_{\xi} \subset I=(\tau_1 , \tau_2)$, and  the limit $\xi \to 0$ should be taken. The normalization constant $C(1, s)$ is defined such that the integral of $h_s$ over $\mathbb{R}$ is normalized to unity
\begin{equation}
\qquad 
C(1, s)=\frac{\Gamma(1 / 2) \sin (\pi s)}{\pi^{3 / 2}}
\, .
\label{fsm-fnp-frac-s-harm-c}
\end{equation}
In our notation,  $s=\alpha/2$. We can easily verify that $h_s (\tau )$ satisfies the equation (\ref{fsm-fnp-frac-Lapl-2-a}) by noting that \cite{Abatangelo:2015} 
\begin{align}
\left( - \Delta_{\tau} \right)^{\frac{\alpha}{2}} h_s (\tau) 
& = C(1,s) \,
\mathrm{P.V.} 
\int_{\mathbb{R} \backslash I_{\xi}(\tau)}
\frac{ h_s(\tau )- h_s(\zeta )}{ |\zeta - \tau|^{1+2s} } d \zeta
\nonumber
\\
& =
C(1,s) \, 
\lim_{\xi \to 0} \int_{\mathbb{R} \backslash I_{\xi}(\tau)}
\frac{ \xi^{2 s} \left(  h_s(\tau )- h_s(\zeta )\right)}{ \vert \zeta - \tau \vert 
\left( |\zeta - \tau|^2 - \xi^2 \right)^s} 
d \zeta = 0
\, .
\label{fsm-fnp-frac-s-harm-0}
\end{align}
If we take $I = (-1,+1)$, the fundamental solution has the following form
\begin{equation}
x^a (\tau ) = (1-\tau^2)^{\frac{\alpha}{2}-1} c^{a}
\, ,
\label{fsm-fnp-frac-Lapl-sol-1}
\end{equation}
where $c^a$ is a constant vector in $\mathbb{R}^n$. This solution can be generalized to standard harmonic functions as follows: If the function $h^a(\tau )$ satisfies the equation $ \Delta_{\tau} h^a (\tau ) = 0$ in $I$, i. e., $h^a(\tau )$ is an harmonic function in $I$, then 
\begin{equation}
x^a (\tau ) = (1-\tau^2)^{\frac{\alpha}{2}-1} h^{a}(\tau )
\, ,
\label{fsm-fnp-frac-Lapl-sol-2}
\end{equation}
satisfies the equation (\ref{fsm-fnp-frac-Lapl-2-a}) in $I$ \cite{Hmissi:1994,Bogdan:1999}.  

To observe the effect of fractionality parameter on free particle motion, let's consider specific values for $\alpha$. When $\alpha =2$, equation (\ref{fsm-fnp-frac-Lapl-sol-2}) yields the trajectory of a Newtonian particle $x^a (\tau ) = h^a (\tau) = v^a \tau + x^a_0$ from $\tau_1$ to $\tau_2$, where $v^a$ and $x^a_0$ represent the components of the constant velocity and initial position vector. This result arises from the known property that the only harmonic polynomial of degree one is a linear polynomial. Note that the initial and final conditions give: $- v^a + x^a_0 = v^a  + x^a_0 = 0 $ which fixes both the velocity and initial position to zero, as expected. However, for $\alpha = 1$, the trajectory of the $1/2$ - fractional particle is given by 
\begin{equation}
x^a (\tau ) = \frac{v^a \tau + x^a_0}{\sqrt{1-\tau^2 }}
\, ,
\label{fsm-fnp-frac-Lapl-sol-3}
\end{equation}
which can be compared with the solution from equation (\ref{fsm-fnp-frac-Lapl-sol-2}) above. As the formula (\ref{fsm-fnp-frac-Lapl-sol-3}) demonstrates, the trajectory of a $1/2$-fractional particle becomes singular as $\tau \to \pm 1$ for arbitrary values of $v^a$ and $x^a_0$. This singularity is characteristic of the fractional Dirichlet problem, indicting information on the singularities of the solutions on the boundary of the parameter set \cite{Chen:2020}. Nevertheless, when $v^a = x^a_0 = 0$ as in the case of a Newtonian particle, the $1/2$-fractional particle obeys the law of inertia with a non-singular and vanishing solution. However, we can see that there are some interesting features of the fractional particle that differ from the standard particle.

Let's discuss the physical interpretation of the trajectory defined by equation (\ref{fsm-fnp-frac-Lapl-sol-3}). Firstly, it's important to note that the trajectory represents a hyperbolic path in $\mathbb{R}^n$. This characteristic is evident from the denominator $\sqrt{1-\tau^2}$, indicating singularities at $\tau = \pm 1$, where the particle reaches infinite distance in proper time. 
Secondly, the initial conditions exert a significant influence on the trajectory. The particle's initial velocity $v^a$ and initial position $x_0^a$ determine the shape and orientation of the hyperbolic path in $\mathbb{R}^n$. Altering the initial velocity would consequently change the slope and direction of the trajectory.
Thirdly, the trajectory described by equation (\ref{fsm-fnp-frac-Lapl-sol-3}) manifests two singularities at $\tau = \pm 1$, indicating unbounded motion of the particle at these points. This could imply that the particle attains infinite distance or velocity at these instances. The presence of the fractional Laplacian operator in the kinetic term suggests the possibility of the particle experiencing anomalous diffusion or motion influenced by long-range interactions.

%-----------------------------------------------------------------
%-----------------------------------------------------------------
\subsection{Fractional Harmonic Oscillator}
\label{sec:exfp-fho}

The fractional harmonic oscillator, characterized by a fractional particle subject to an external potential, introduces a new level of complexity to the oscillator paradigm. The fractionality can be realized through a diverse array of fractional time derivatives, each giving rise to an unique model with physical applications outlined in the introduction. The precise correlation between the physical attributes of the systems in question and the fractional nature of the mathematical models remains an active area of research and interpretation \cite{Stanislavsky:2004}. Beside the examples mention in the introduction, the fractional harmonic operators find their utility in various domains of high-energy physics, such as the resolution of the fractional Schrödinger equation \cite{Herrmann:2018}, quantum gravity \cite{Giesel:2021wsy}, fractional supersymmetric quantum mechanics where the Hamiltonian is proportional to the $p$-th power of the supercharge operator \cite{Quesne:2002hj,Daoud:2003qy}, and path integral quantization \cite{Eab:2006}. In this subsection, we discuss the fractional harmonic oscillator equipped with a fractional Laplacian. 

The fractional harmonic oscillator characterizes a fractional particle subject to an external potential defined as
\begin{equation}
V[x] = \frac{|k|}{2} x_a  x^a 
\, ,
\label{fsm-fnp-frac-osc-1}
\end{equation}
where $\tau \in \mathbb{R}$. The equation of motion for a fractional harmonic oscillator, derived from (\ref{fsm-fnp-L-frac-eom-3}), 
resembles the eigenvectors and eigenvalues equation for the fractional Laplacian operator
\begin{equation}
\left(-\Delta_{\tau} \right)^{\frac{\alpha}{2}} x^{a} (\tau) 
= - \frac{|k|}{m_{\alpha}}  x^a (\tau) 
\, .
\label{fsm-fnp-P-frac-osc-2}
\end{equation}
Solving (\ref{fsm-fnp-P-frac-osc-2}) leads to expressions in terms of fractional harmonic exponentials. Using definition (\ref{fsm-fnp-L-frac-def}) and after a simple algebra, we arrive at the solution
\begin{equation} 
x^{a (\pm )}_{\omega, \alpha} (\tau ) = x^a_0 \exp \left[ \pm i \left( \frac{|k| }{m_{\alpha}}\right)^{\frac{1}{\alpha}} \tau \right]
\, .
\label{fsm-fnp-P-frac-eom-osc-3}
\end{equation}
The functions $x^{a (\pm )}_{\omega, \alpha} (\tau )$ show that fractional particle oscillates with frequency determined by 
\begin{equation}
\omega_{\alpha}  = \left( \frac{|k|}{m_{\alpha}} \right)^{\frac{1}{\alpha}}
\, .
\label{fsm-fnp-osc-frac-eom-osc-4}
\end{equation}
In the case of $\alpha=2$, we recover the standard harmonic oscillator. 

The trajectory of the fractional particle, as described by equation (\ref{fsm-fnp-P-frac-eom-osc-3}), is periodic in all directions of the target-space. The period, denoted by $T_{\alpha}$, is given by $2 \pi/ \omega_{\alpha}$. Let's consider a time interval $[ \tau_1 , \tau_2 = \tau_1 + T_{\alpha} ]$. We impose initial and final conditions on $x^{a} (\tau )$ such that it vanishes at the boundary points $\{ \tau_1, \tau_2 \}$, as shown in equation  
\begin{equation}
x^{a}(\tau_1) = x^{a}(\tau_2 ) = 0
\, .
\label{fsm-fnp-osc-frac-eom-osc-5}
\end{equation}
Equations (\ref{fsm-fnp-P-frac-osc-2}) and (\ref{fsm-fnp-osc-frac-eom-osc-5}) together define a time-like Dirichlet problem for the fractional Laplacian. It is important to note that the real and imaginary parts of $x^{a (\pm )}_{\omega, \alpha} (\tau )$ satisfy equation (\ref{fsm-fnp-P-frac-osc-2}) independently due to the linearity of the fractional Laplacian. The general solution of the equation (\ref{fsm-fnp-P-frac-osc-2}) is a Fourier sum of cosine and sine functions. Consequently, the general solution of the Dirichlet problem, defined by equations (\ref{fsm-fnp-P-frac-osc-2}) and (\ref{fsm-fnp-osc-frac-eom-osc-5}), can be expressed as a sum of sine functions only
\begin{equation}
x^{a}_{\alpha} (\tau ) = - \sum_{l=1}^{\infty} x^{a}_{\alpha , l} (\tau ) = 
- \sum_{l=1}^{\infty} B^{a}_{\alpha , l} \left( \frac{2 \pi l}{\tau_2 - \tau_1} \right)^{\alpha}
\sin \left[ \frac{2 \pi l(\tau - \tau_1)}{\tau_2 - \tau_1} \right]
\, .
\label{fsm-fnp-osc-frac-5}
\end{equation}
The coefficients $B^{a}_{l}$ in equation (\ref{fsm-fnp-osc-frac-5}) can be determined by imposing the normalizing condition
\begin{equation}
\int_{\tau_1}^{\tau_1 + T_{\alpha}} d \tau \, x^{a}_{\alpha, l} (\tau ) \, x^{b}_{\alpha , r} (\tau) = \delta^{a b} \delta_{l r}
\, .
\label{fsm-fnp-osc-frac-6}
\end{equation}
After performing some calculations, we find that the normalized modes are given by 
\begin{equation}
x^{a}_{\alpha , l} (\tau )= \pi^{-1}
\left( \frac{|k|}{m_{\alpha}} \right)^{\frac{1}{2 \alpha}} 
\sin \left[ l \left( \frac{|k|}{m_{\alpha}} \right)^{\frac{1}{ \alpha}} \left( \tau - \tau_1 \right) \right] u^a
\, ,
\label{fsm-fnp-osc-frac-7}
\end{equation}
where $u^a$ is the unit vector in the $a$-direction. We can calculate the instantaneous velocity of the fractional oscillator, $v^{a}_{\alpha} (\tau )$, by taking the first derivative of $x^{a}_{\alpha} (\tau )$ with respect to $\tau$. The result is given in the following equation
\begin{equation}
v^{a}_{\alpha} (\tau ) = - \pi^{-1} \left( \frac{|k|}{m_{\alpha}} \right)^{\frac{3}{2\alpha}}
\sum_{l=1}^{\infty} l
\cos \left[ \left( \frac{|k|}{m_{\alpha}} \right)^{\frac{1}{2\alpha}} \left(\tau - \tau_1 \right)\right]
u^{a}
\, .
\label{fsm-fnp-osc-frac-8}
\end{equation}

As discussed in the previous section, the kinetic energy is the product of $\mathbf{v}_{\alpha}$ and $\bm{\pi}^{\alpha}$. Given that the fractional oscillating modes are independent, the kinetic energy can be expressed as a sum over modes, as shown in the following equation 
\begin{equation}
E^{(\alpha)} = \sum_{l = 1}^{\infty} E^{(\alpha)}_{l}
\, .
\label{fsm-fnp-osc-frac-9}
\end{equation}
To determine the instantaneous energy of each mode $E^{(\alpha)}_{l}$, we use equations (\ref{fsm-fnp-can-mom}) and (\ref{fsm-fnp-kin-energ}). The canonical momentum $\pi^{a}_{\alpha, l}$ is obtained by substituting equation (\ref{fsm-fnp-osc-frac-7}) into equation (\ref{fsm-fnp-can-mom}). This leads to the calculation of the action of $\left(- \Delta^{1}_{\tau} \right)^{\frac{\alpha}{2}}$ on the modes $x^{a}_{\alpha , l} (\tau )$. 
Let us determine the instantaneous energy of each mode $E^{(\alpha)}_{l}$. To this end, we use the equations (\ref{fsm-fnp-can-mom}) and (\ref{fsm-fnp-kin-energ}) and observe that the canonical momentum $\pi^{a}_{\alpha, l}$ is obtained by substituting the equation (\ref{fsm-fnp-osc-frac-7}) into the equation (\ref{fsm-fnp-can-mom}) which leads to the calculation of the action of $\left(- \Delta^{1}_{\tau} \right)^{\frac{\alpha}{2}}$ on the modes $x^{a}_{\alpha , l} (\tau )$. The result can be found by transforming the first-order fractional derivative into the integral 
\begin{equation}
\left(- \Delta^{(1)}_{\tau} \right)^{\frac{\alpha}{2}} x^{a}_{\alpha , l} (\tau ) =
2^{\alpha - 1} \pi^{-\frac{3}{2}} \left( \frac{|k|}{m_{\alpha}} \right)^{\frac{1}{2\alpha}}
\frac{\Gamma \left( \frac{1+\alpha}{2} \right) }{\Gamma\left( 1- \frac{\alpha}{2} \right)}
\int_{-\infty}^{+\infty} d \zeta \frac{\zeta}{|\zeta |^{1+\alpha}}
\sin \left[ l \left( \frac{|k|}{m_{\alpha}} \right)^{\frac{1}{2\alpha}} \left( \zeta + \tau - \tau_1 \right) \right]
u^a
\, .
\label{fsm-fnp-osc-frac-10}
\end{equation} 
After some algebraic manipulations, we obtain the following equation 
\begin{align}
\left(- \Delta^{(1)}_{\tau} \right)^{\frac{\alpha}{2}} x^{a}_{\alpha , l} (\tau ) 
& =
2^{\alpha} \pi^{-\frac{3}{2}} \left( \frac{|k|}{m_{\alpha}} \right)^{\frac{(\alpha - 1)^2 + \alpha^2 }{2 \alpha}}
\frac{\Gamma \left( \frac{1+\alpha}{2} \right) \Gamma \left( 1- \alpha \right) }{\Gamma\left( 1- \frac{\alpha}{2} \right)}
\nonumber
\\
& \times \sin \left[ \frac{(1-\alpha) \pi}{2} \right]
\cos \left[ l \left( \frac{|k|}{m_{\alpha}} \right)^{\frac{1}{2\alpha}} \left( \zeta + \tau - \tau_1 \right) \right]
u^a
\, ,
\label{fsm-fnp-osc-frac-11}
\end{align}
which holds for all $0 < \alpha < 2$. From this equation, we can calculate the canonical momentum $\pi^{a}_{\alpha , l}$ corresponding to the $l$-th mode
\begin{align}
\pi^{a}_{\alpha , l}
& =
- 2^{\alpha - 1} \pi^{-\frac{3}{2}} m_{\alpha} \left( \frac{|k|}{m_{\alpha}} \right)^{\frac{(\alpha - 1)^2 + \alpha^2 }{2 \alpha}}
\frac{\Gamma \left( \frac{1+\alpha}{2} \right) \Gamma \left( 1- \alpha \right) }{\Gamma\left( 1- \frac{\alpha}{2} \right)}
\nonumber
\\
& \times \sin \left[ \frac{(1-\alpha) \pi}{2} \right]
\cos \left[ l \left( \frac{|k|}{m_{\alpha}} \right)^{\frac{1}{2\alpha}} \left( \zeta + \tau - \tau_1 \right) \right]
u^a
\, .
\label{fsm-fnp-osc-frac-12}
\end{align}
The instantaneous kinetic energy of the $l$-th mode can be obtained from equation (\ref{fsm-fnp-kin-energ}) and equations (\ref{fsm-fnp-osc-frac-8}) and (\ref{fsm-fnp-osc-frac-12}). The result has the following form
\begin{align}
E^{\alpha}_{l} (\tau ) 
& = 
2^{\alpha - 2} \pi^{- \frac{5}{2}} n l m_\alpha \left( \frac{|k|}{m_{\alpha}} \right)^{\frac{(\alpha - 1)^2 + \alpha^2 + 3 }{2 \alpha}} 
\frac{\Gamma \left( \frac{1+\alpha}{2} \right) \Gamma \left( 1- \alpha \right) }{\Gamma\left( 1- \frac{\alpha}{2} \right)}
\nonumber
\\
& \times \sin \left[ \frac{(\alpha - 1) \pi}{2} \right]
\cos^2 \left[ l \left( \frac{|k|}{m_{\alpha}} \right)^{\frac{1}{2\alpha}} \left( \zeta + \tau - \tau_1 \right) \right]
\, .
\label{fsm-fnp-osc-frac-13}
\end{align}

The equation above shows that $E^{\alpha}_{l} (\tau )$ depends on the space dimension $n$, as it accounts for all degrees of freedom along all space directions. The particle's instantaneous kinetic energy, given by the sum on the right-hand side of equation (\ref{fsm-fnp-osc-frac-9}), is divergent. However, a more meaningful observable is the average kinetic energy $\bar{E}^{\alpha}$ over the period $T_\alpha$. This is the sum of the average energy of the $l$-modes, and it is given by the following equation
\begin{equation}
\bar{E}^{\alpha}_{l} = \frac{1}{T_\alpha} \int_{\tau_1}^{\tau_1 + T_\alpha} d \tau E^{\alpha}_{l} (\tau )
\, .
\label{fsm-fnp-osc-frac-14}
\end{equation}
Note that the only $\tau$-dependent factor in equation (\ref{fsm-fnp-osc-frac-13}) is the last one. According to equation (\ref{fsm-fnp-osc-frac-eom-osc-4}), this is equal to $\cos^2(2 \pi l (\tau - \tau_1)/T_{\alpha})$, whose average over the period $T_{\alpha}$ is one-half. Therefore, the average energy $\bar{E}^{\alpha}_{l}$ is proportional to $\sum_{l=1}^{\infty} l$. By applying any regularization formula for the sum, such as the Dirichlet series regularization 
\begin{equation}
\lim_{s \to 0 } \left(  \sum_{l=1}^{\infty} l^{1-s} \right) = - \frac{1}{12}
\, ,
\label{fsm-fnp-osc-frac-14-1}
\end{equation}
we obtain the regularized value for the average energy
\begin{equation}
\bar{E}^{\alpha} 
= 
\frac{2^{\alpha - 5}  n l m_\alpha}{\pi^{\frac{5}{2}}3} 
\left( \frac{|k|}{m_{\alpha}} \right)^{\frac{(\alpha - 1)^2 + \alpha^2 + 3 }{2 \alpha}} 
\frac{\Gamma \left( \frac{1+\alpha}{2} \right) \Gamma \left( 1- \alpha \right) }{\Gamma\left( 1- \frac{\alpha}{2} \right)}
\sin \left[ \frac{(1- \alpha) \pi}{2} \right]
\, .
\label{fsm-fnp-osc-frac-15}
\end{equation}

The instantaneous potential of a particle moving according to the solution 
\begin{equation}
x^{a}_{\alpha} (\tau )= - \sum_{l=1}^{\infty} \pi^{-1}
\left( \frac{|k|}{m_{\alpha}} \right)^{\frac{1}{2 \alpha}} 
\sin \left[ l \left( \frac{|k|}{m_{\alpha}} \right)^{\frac{1}{ \alpha}} \left( \tau - \tau_1 \right) \right] u^a
\, ,
\label{fsm-fnp-osc-frac-16}
\end{equation} 
can be easily calculated, yielding the following formula 
\begin{equation}
V\left[ x_{\alpha} (\tau ) \right] =
\frac{n |k|}{2 \pi^2}
\left( \frac{|k|}{m_{\alpha}} \right)^{\frac{1}{\alpha}}
\sum_{l=1}^{\infty} 
\sin^2 \left[ l \left( \frac{|k|}{m_{\alpha}} \right)^{\frac{1}{ \alpha}} \left( \tau - \tau_1 \right) \right]
\, .
\label{fsm-fnp-osc-frac-17}
\end{equation}
The regularized divergent sum on the right-hand side of equation (\ref{fsm-fnp-osc-frac-17}) vanishes. However, it is possible to calculate the average potential energy, which is also divergent, but whose regularized expression is 
\begin{equation}
\bar{V}\left[ x_{\alpha} (\tau ) \right] = -
\frac{n |k|}{8 \pi^2}
\left( \frac{|k|}{m_{\alpha}} \right)^{\frac{1}{\alpha}}
\, .
\label{fsm-fnp-osc-frac-18}
\end{equation}
This concludes the computation of the primary classical physical observables associated with the fractional harmonic oscillator in $n$ dimensions. Our results are valid for $0 < \alpha < 2$, as previously mentioned. Outside this range, the integrals defining the canonical momentum lack analytic solutions. Consequently, our discussion excludes the classical Newtonian particle defined by $\alpha=2$. In this case, this value must be fixed from the outset, as noted in the previous section. This approach reproduces the well-known results of the classical harmonic oscillator.

The results concerning the fractional harmonic oscillator stem from the solvability of the Dirichlet problem associated with the fractional Laplacian. This solvability property might not extend to other fractional integral or differential operators utilized in alternative harmonic oscillator formulations, or even for models involving the fractional Laplacian. Furthermore, it is our ability to regularize the average energy that permits the derivation of an explicit analytical expression for the oscillator's mean energy.

%-----------------------------------------------------------------
%-----------------------------------------------------------------
\subsection{Charged Fractional Particle in Classical Electromagnetic Field}
\label{sec:exfp-em}

As a last example, let us examine the influence of the electromagnetic field on a fractional particle. There are several possibilities to generalize the interaction between a charged particle and the electromagnetic field to the fractional case. Here, we consider the simplest case of a local interaction between the particle of charge $q$ and the electromagnetic potentials $\phi (\tau, \mathbf{x} )$ and $\mathbf{A} (\tau, \mathbf{x} )$ in $d=3$ dimensions. Given that this is the conventional classical interaction, only the kinetic term of the Lagrangian is altered, as per the relation (\ref{fsm-fnp-act-newt-01}). Consequently, the action functional is expressed as
\begin{equation}
S^{(\alpha)} [x, \mathbf{A}, \phi ] =  \int_{-\infty}^{+\infty} d\tau \, 
\left[
- \frac{m_{\alpha}}{2}
x^a (\tau) \left( - \Delta_{\tau} \right)^{\frac{\alpha}{2}} x^b (\tau ) \eta_{ab} 
+ q \frac{x^a (\tau)}{d \tau } A^{b}(x,\tau) \eta_{ab} - q \phi (x,\tau)
\right]
\, ,
\label{fp-em-action}
\end{equation}
where $a, b,\ldots = 1, 2, 3$. The equation of motion derived from $S^{(\alpha)} [x, \mathbf{A}, \phi ]$ has the following form
\begin{equation}
m_{\alpha} \left( - \Delta_{\tau} \right)^{\frac{\alpha}{2}} x_{a} (\tau ) - q F_{ab} (\tau , \mathbf{x} ) \frac{d x^{b} (\tau)}{d \tau} + \partial_a \phi (\tau , \mathbf{x} ) + \partial_{\tau} A_{a} (\tau , \mathbf{x} ) = 0
\, ,
\label{fp-em-eom}
\end{equation}
where $F_{ab} = \partial_{a} A_{b} - \partial_{b} A_{a}$. Assuming that $x^a (\tau)$ remains unchanged under the gauge transformations
\begin{equation}
\mathbf{A} \rightarrow \mathbf{A}^{\prime} = \mathbf{A} + \bm{\nabla} \Lambda
\, ,
\qquad
\phi \rightarrow \phi^{\prime} = \phi - \partial_{\tau} \Lambda
\, ,
\label{fp-em-gauge-tr}
\end{equation}
the Lagrangian $L^{\alpha} [x; \mathbf{A} , \phi]$ undergoes a transformation similar to that in classical mechanics:
\begin{equation}
L^{\alpha} [x; \mathbf{A}^{\prime} , \phi^{\prime}] = 
L^{\alpha} [x; \mathbf{A} , \phi] + \frac{d \Lambda}{d \tau}
\, ,
\label{fp-em-gauge-lag}
\end{equation}
where $\Lambda$ is the gauge parameter. 

The transformation (\ref{fp-em-gauge-lag}) arises from the similarity of the interaction term in $L^{\alpha} [x; \mathbf{A} , \phi]$ to that in classical mechanics. The equation of motion (\ref{fp-em-eom}) is a hybrid differential equation, combining the fractional Laplacian and the standard first derivative in the $\tau$ variable. Expressing the electric and magnetic fields in terms of potentials, i.e., $\mathbf{E} = - \bm{\nabla} \phi - \partial_{\tau} \mathbf{A}$ and $\mathbf{B} = \bm{\nabla} \times \mathbf{A}$, allows us to rewrite the equations of motion in terms of the electromagnetic field as follows:

\begin{equation}
m_{\alpha} \left( - \Delta_{\tau} \right)^{\frac{\alpha}{2}} x_{a} (\tau ) = 
q\left[ E_a (\tau , \mathbf{x} )+ ( \frac{d \mathbf{x}}{d \tau} \times \mathbf{B} (\tau, \mathbf{x} ) )_{a} \right]
\, .
\label{fp-em-eom-Lorentz}
\end{equation}
The right-hand side of the equation above exhibits the Lorentz force, where the instantaneous velocity plays the same role as in classical electrodynamics. This indicates that, under the assumption of a local interaction between the fractional particle and the classical field, the dynamics of the fractional particle is governed by the same electromagnetic force as that of a standard particle. However, notable distinctions emerge due to the presence of the fractional Laplacian. To elucidate these differences, let us take a closer look at the specific scenario of a charged fractional particle subjected to a constant magnetic field
\begin{equation}
\mathbf{E} = \bm{0}
\, ,
\qquad
\mathbf{B} = (0,0,B)
\, ,
\label{fp-em-const-B-1}
\end{equation}
where $B=\text{constant}$ denotes the field component along the vertical axis.

The equation of motion (\ref{fp-em-eom-Lorentz}) takes the following form
\begin{align}
\left( - \Delta_{\tau} \right)^{\frac{\alpha}{2}} x_{1} (\tau )
& = 
\omega^{\alpha}_{c} \frac{d x_{2} (\tau ) }{d \tau} 
\, ,
\label{fp-em-const-B-2}
\\
\left( - \Delta_{\tau} \right)^{\frac{\alpha}{2}} x_{2} (\tau )
& = 
- \omega^{\alpha}_{c} \frac{d x_{1} (\tau ) }{d \tau} 
\, ,
\label{fp-em-const-B-3}
\\
\left( - \Delta_{\tau} \right)^{\frac{\alpha}{2}} x_{3} (\tau )
& = 0
\, ,
\label{fp-em-const-B-4}
\end{align}
where we have introduced the fractional equivalent of the cyclotron frequency
\begin{equation}
\omega^{\alpha}_{c} = \frac{q B}{m_{\alpha}}
\, .
\label{fp-em-const-B-5}
\end{equation}
Since the motion along the vertical axis is governed by the free fractional particle equation of motion (\ref{fp-em-const-B-4}), as was discussed in the subsection \ref{sec:exfp-free}, our attention now shifts to examining the first two equations (\ref{fp-em-const-B-2}) and (\ref{fp-em-const-B-3}). These equations describe the trajectory of the fractional charged particle within the horizontal plane.

We can solve equations (\ref{fp-em-const-B-2}) and (\ref{fp-em-const-B-3}) by recalling that the exponential function is an eigenfunction of the operator $\left( - \Delta_{\tau} \right)^{\frac{\alpha}{2}} $. Introducing the complex functions
\begin{equation}
z^{\alpha} (\tau ) = x^{\alpha}_{1} (\tau ) + i x^{\alpha}_{2} (\tau )
\, ,
\label{fp-em-const-B-6}
\end{equation}
equations (\ref{fp-em-const-B-1}) and (\ref{fp-em-const-B-2}) can be equivalently expressed as the complex equation
\begin{equation}
\left( - \Delta_{\tau} \right)^{\frac{\alpha}{2}} z^{\alpha} (\tau )
= 
-i \omega^{\alpha}_{c} \frac{d z^{\alpha} (\tau ) }{d \tau} 
\label{fp-em-const-B-7}
\end{equation}
A comparison of the equations (\ref{fsm-fnp-P-frac-osc-2}) and (\ref{fp-em-const-B-7}) suggests the ansatz
\begin{equation}
z^{\alpha} (\tau ) = Z^{\alpha} e^{i \omega^{\prime}_{\alpha} \tau}
\, ,
\label{fp-em-const-B-8}
\end{equation}
where $Z^{\alpha} \in \mathbb{C}$ is a constant and $\omega^{\prime}_{\alpha}$ denotes the oscillating frequency of the fractional particle in the horizontal plane. Substituting the right-hand side of the equation (\ref{fp-em-const-B-8}) into the equation (\ref{fp-em-const-B-7}) we obtain the relation
\begin{equation}
\omega^{\prime}_{\alpha} = \left( \frac{q B}{m_{\alpha}} \right)^{\frac{1}{\alpha - 1}}
\, .
\label{fp-em-const-B-9}
\end{equation}
Without loss of generality, we can set the phase of $Z^{\alpha}$ such that it is real. Consequently, the motion of the fractional particle in the horizontal plane is described by the following functions
\begin{align}
x^{\alpha}_{1} (\tau ) = \mathit{x}^{\alpha}_{1} + Z^{\alpha} \cos \left[ \left( \frac{q B}{m_{\alpha}} \right)^{\frac{1}{\alpha - 1}} \tau \right]
\, ,
\label{fp-em-const-B-10}
\\
x^{\alpha}_{1} (\tau ) = \mathit{x}^{\alpha}_{2} + Z^{\alpha} \sin \left[ \left( \frac{q B}{m_{\alpha}} \right)^{\frac{1}{\alpha - 1}} \tau \right]
\, , 
\label{fp-em-const-B-11}
\end{align}
where $\mathit{x}^{\alpha}_{1}$ and $\mathit{x}^{\alpha}_{2}$ are real constants.

Let us conclude this section with some important remarks. As briefly discussed earlier, the boundary conditions, which in this case are initial and final conditions, wield considerable influence over the solutions to the equations of motion. Confining the particle's motion within a specific  open time interval, these conditions not only determine the constant factors but also dictate the nature of solutions, often encoding information about temporal discontinuities at the trajectory's initial and final instants. Such behaviour typifies fractional Laplacians in either one or more dimensions \cite{Chen:2020}.

The models developed herein extend classical Newtonian frameworks directly, by fractionalizing solely the kinetic term and assuming local interactions. 
However, other models can be constructed by incorporating non-local interactions between particles and fields. For instance, one could explore fractional electromagnetic fields as investigated in \cite{LaNave:2019mwv,Porto:2023,Heydeman:2022yni}. An in-depth analysis of such models, while pertinent, lies beyond the scope of this paper.

%-----------------------------------------------------------------
\section{Sigma Model from Free Fractional Particle}
\label{sec:fsm-free}

Among the diverse class of fractional particle models, which arise from the formulation of kinetic and potential terms utilizing fractional operators, the free fractional classical particle discussed in the previous section stands out for its intriguing property that it can be mapped into a local classical sigma model by extending the world-line to the superior half-plane and imposing appropriate boundary conditions. The well-established Caffarelli-Silvestre extension problem \cite{Caffarelli:2007} ascertain that this mapping is well defined. This result has been used to establish a similar correspondence between a non-local fractional scalar field in $m$-dimensions and a local scalar field in $m+1$-dimensions. 
Some properties of these non-local and local scalar fields have been explored in \cite{Rajabpour:2011qr,Paulos:2015jfa,Frassino:2019yip}. 

%-----------------------------------------------------------------
%-----------------------------------------------------------------
\subsection{Classical Sigma Model}
\label{sec:fsm-free-class}

In order to map the classical fractional particle to the local sigma model,  we associate a field $X^a : \mathbb{R} \times \mathbb{R}_{+} \rightarrow \mathbb{R}$ which is a continuous functions in its arguments 
to each field  $x^a: \mathbb{R} \rightarrow \mathbb{R}$, and we require that the following equations be satisfied
\begin{align} 
X^a (\tau, 0) & = x^a (\tau) 
\, ,
\label{fsm-flp-ext-1}
\\ 
\partial_{\mu} \left[ \sigma^{1-\alpha} \partial^{\mu} X^a (\tau,\sigma) \right]
& = 0
\, ,
\label{fsm-flp-ext-2}
\end{align}
where $\sigma \in \mathbb{R}_{+}$
\footnote{
At times, we will adopt the alternative notation for the two-dimensional coordinates $\sigma^{\mu} = (\sigma^0 , \sigma^1 ) = ( \tau, \sigma )$, primarily for succinctly expressing the action. However, to maintain clarity and avoid redundant upper indices, we will represent $d^2 \sigma (\sigma^{1})^{1 - \alpha}$ as $d \tau d \sigma \sigma^{1-\alpha}$, and so forth.}. By the extension problem of Caffarelli and Silvestre \cite{Caffarelli:2007}, the following equality holds
\begin{equation}
(-\Delta_{\tau})^{\frac{\alpha}{2}} x^a (\tau)= \kappa_\alpha \lim_{\sigma \to 0^{+}} \left[ \sigma^{1-\alpha} \partial_{\sigma} X^a(\tau,\sigma) \right]
\, , 
\label{fsm-flp-ext-3}
\end{equation}
where $\kappa_{\alpha}$ is a positive constant given by
\begin{equation}
\kappa_{\alpha} = \frac{2^{\alpha} \Gamma \left( \frac{\alpha}{2}\right)}{\alpha \Gamma \left( - \frac{\alpha}{2} \right)}
\, .
\label{fsm-fnp-act-frac-C}
\end{equation}
The extension problem given by the equations (\ref{fsm-flp-ext-1})-(\ref{fsm-flp-ext-3}) generalizes the harmonic extension problem for the square root Laplacian $(-\Delta_{\tau})^{-1/2}$ \cite{Caffarelli:2007} to $\alpha \in (0, 2]$. 

The sigma model action is defined as
\begin{equation}
S^{(\alpha )}_0 [X] = \frac{T_{\alpha} \kappa_{\alpha} c }{2} \int_{-\infty}^{+\infty} d\tau \int_{0}^{+\infty} d \sigma 
\sigma^{1-\alpha} \partial_{\mu} X^a (\tau , \sigma) \partial_{\nu } X^b (\tau , \sigma ) \eta^{\mu \nu} \eta_{ab}
\, ,
\label{fsm-fnp-act-def-00}
\end{equation}
where $\partial_{\mu} = \partial/\partial \sigma^{\mu}$ and $\eta_{\mu \nu}$ is the Euclidean metric tensor in the half-plane. The fields $X^{a} (\tau , \sigma )$ represent a map from the half-plane $\mathbb{R} \times \mathbb{R}_{+}$ to the $n$-dimensional space. According to the extension problem (\ref{fsm-flp-ext-3}), the dimension of $\langle X^{a}(\tau , \sigma ) \rangle = \langle x^a(\tau ) \rangle$ which implies that $\langle T_{\alpha} \rangle = E^2 \langle m_{\alpha} \rangle = E^{5-\alpha}$. In what follows, we are going to take $T_{\alpha} = 1$ for simplicity.
Since the fractionality parameter $\alpha$ appears only in the exponent of the $\sigma$ factor, the sigma model is not fractional, that is, it does not contain any fractional operator. An important property guaranteed by the extension problem is that
\begin{equation}
S^{(\alpha)}_0 [x] = S^{( \alpha )}_0 [X]
\, .
\label{fsm-fnp-act-eq}
\end{equation}
After a Wick rotation, $\tau \rightarrow \tau^{\prime} = i \tau$, the action $S^{( \alpha )}_0 [X]$ describes a two-dimensional sigma model in the half Minkowski space which has on its boundary a free fractional particle. However, the Wick rotation changes the sign of the kinetic term of the sigma model which can be made positive by considering a minus sign in the right hand of the equation (\ref{fsm-fnp-act-eq}). 

The equations of motion of $X^a (\tau , \sigma )$ are obtained by applying
the variational principle to $S^{(\alpha )}_0 [X]$, and they are given by
\begin{equation}
\partial_{\tau}^{2} X^{a} (\tau , \sigma) - \frac{1-\alpha }{\sigma} \partial_{\sigma} X^{a} (\tau , \sigma ) - \partial^{2}_{\sigma} X^{a} (\tau , \sigma) = 0
\, .
\label{fsm-fnp-act-eom}
\end{equation}
The boundary conditions to be imposed on $X^{a} (\tau , \sigma )$ follow from the equations (\ref{fsm-flp-ext-1}) and (\ref{fsm-flp-ext-2}), yielding the following set of equations
\begin{equation}
\lim_{\sigma \rightarrow 0} X^{a}(\tau, \sigma)=0 \, , 
\qquad 
\lim_{\sigma \rightarrow 0} 
\left[
\sigma^{1-\alpha} \partial_{\sigma} X^{a}(\tau, \sigma)
\right]
=0
\, .
\label{fsm-fnp-eom-bc}
\end{equation}
It is noteworthy that equations (\ref{fsm-fnp-act-eom}) and (\ref{fsm-fnp-eom-bc}) possess an universal nature. Specifically, any mapping from a fractional field theory in $m$ variables to a local field theory in $m+1$ variables, as established by the extension problem \cite{Caffarelli:2007}, will generate these equations along the $(m+1)$-th direction. In the context of this discussion, $m=1$ and the $(m+1)$-th direction corresponds to the $\sigma$ direction, spanning the range $\mathbb{R}_{+}$.

The equation (\ref{fsm-fnp-act-eom}) can be solved by elementary methods, the following general solution being found 
\begin{equation}
X^a(\tau, \sigma;\alpha) = 
\int_0^{+\infty} \frac{d \mu(\omega)}{\sqrt{2 \omega }}
\left[ 
A^a(\omega) e^{-i\omega \tau } \varphi(\sigma, \omega;\alpha)
+ B^a(\omega) 
e^{i\omega \tau } \bar{\varphi}(\sigma, \omega;\alpha)\right]
\, .
\label{fsm-fnp-eom-sol}
\end{equation}
In order to avoid the proliferation of indices, we will drop $\alpha$ from the arguments of $X^a$ and $\varphi$. The reality condition $X^a(\tau, \sigma ) = \bar{X}^{a}(\tau , \sigma )$ imposes the reality of the coefficients $\bar{A}^a (\omega) = B^a(\omega) $. The main properties of the measure $d \mu (\omega )$ and the functions $\varphi (\sigma, \omega )$ were obtained in \cite{Frassino:2019yip}. Specifically, it was demonstrated that
\begin{align}
d \mu(\omega) 
& = 
2^{\alpha} \pi^{-2} \Gamma^2\left(\frac{\alpha}{2}\right) 
\sin^2 \left( \frac{\pi \alpha}{2}\right)
\omega^{1-\alpha} d \omega 
\nonumber
\\
& = 
\mu (\alpha ) \omega^{1-\alpha} d \omega =
\mu (\omega ) d \omega
\, ,
\label{fsm-fnp-eom-mu}
\\
\varphi(\sigma, \omega) 
& = 
\frac{\pi (\omega \sigma)^{\frac{\alpha}{2}} J_{-\frac{\alpha}{2}}(\omega \sigma)}{2^{\frac{\alpha}{2}} \Gamma\left(\frac{\alpha}{2}\right) \sin \left(\pi \frac{\alpha}{2}\right)}
\, .
\label{fsm-fnp-eom-phi}
\end{align} 
The functions $\varphi (\sigma, \omega )$ satisfy the following orthogonality conditions
\begin{align}
\int_0^{+\infty} d \mu(\omega ) 
\bar{\varphi}
\left(
\omega, \sigma \right) \varphi \left(\omega, \sigma^{\prime} \right)
& =
\sigma^{\alpha-1} \delta \left(\sigma-\sigma^{\prime} \right)
\, .
\label{fsm-fnp-eom-ort}
\\
\int_0^{+\infty} d \mu(\sigma ) 
\bar{\varphi}
\left(
\omega, \sigma \right) \varphi \left(\omega^{\prime}, \sigma \right)
& =
\omega^{\alpha-1} \delta \left(\omega - \omega^{\prime} \right)
\, .
\label{fsm-fnp-eom-ort-1}
\end{align}
The equation (\ref{fsm-fnp-eom-ort-1}) can be obtained from (\ref{fsm-fnp-eom-ort}) by observing that $\varphi (\sigma , \omega )$ is symmetric in its variables. If we use the common convention that the determination of the $\nu$-th power argument $z$ of the Bessel function $J_{\nu} (z)$ determines the reality of the function for $z \in \mathbb{R}$, then $J_{\nu} (x)$ is real for real $\nu$ and $x \in \mathbb{R}_{+}$, which is the case described by the equation (\ref{fsm-fnp-eom-phi}) above, where $\nu = - \alpha/2$ and $x=\omega \sigma$ \cite{Abramowitz:1972}. With this convention, the boundary limit on the right-hand side of the equation (\ref{fsm-fnp-eom-sol}) is given by
\begin{equation}
x^{a} (\tau ) = 
\lim_{\sigma \to 0} X^a(\tau, \sigma) = 
D_{\alpha}
\int_0^{+\infty} d \mu(\omega) \,
\omega^{\frac{\alpha-1}{2}} 
\left[ 
A^a(\omega) e^{-i\omega \tau } 
+ B^a(\omega) 
e^{i\omega \tau } 
\right]
\, ,
\label{fsm-fnp-eom-sol-limit}
\end{equation}
where
\begin{equation}
D_{\alpha} = 
\frac{1}{
2^{\frac{3}{2}}
\sin \left(\frac{\alpha \pi}{2} \right)
\Gamma\left( \frac{\alpha}{2} \right)
\Gamma\left(1-\frac{a}{2}\right)}
\, .
\label{fsm-fnp-sol-const-D}
\end{equation}
The last two equations establish a concrete correspondence between the solutions of the classical sigma model given by the equation 
(\ref{fsm-fnp-eom-sol}) and the fractional particle coordinates on the boundary at $\sigma =0$. 

%-----------------------------------------------------------------
%-----------------------------------------------------------------
\subsection{Symmetries of Sigma Model}
\label{sec:fsm-symm}

In this subsection, we analyse the symmetries of the sigma model defined by the Caffarelli-Silvestre extension problem and explore the conserved quantities of the theory. While the analytical procedure remains consistent for both $SO(n)$ and $SO(n,1)$ groups, we choose to present the latter for generality. The results from this section can be directly applied to corresponding situations in Euclidean space and the Newtonian fractional particle model discussed in this work. According to the Noether's theorem, the invariance of $S^{(\alpha )}_0 [X]$ under global symmetries corresponds to conserved charged, while the presence of local symmetries indicates a redundancy of the degrees of freedom. 

From equation (\ref{fsm-fnp-act-def-00}), it is evident that $S^{(\alpha )}_0 [X]$ is invariant under the Poincaré transformations in $n+1$-dimensions
\begin{equation}
\delta X^{a} (\sigma^{\mu} ) = {\Lambda^{a}}_{b} X^{b} (\sigma^{\mu} ) + \xi^{a}
\, ,
\label{fsm-fnp-symm-1}
\end{equation}
where $\Lambda \in SO(n,1)$ and $\xi$ is a constant $n+1$-dimensional vector. According to Noether's theorem, there exist conserved currents associated with the Poincaré symmetry. To determine these currents, we apply Noether's method wherein the parameters of the Poincaré transformations are made functions on $\sigma^{\mu}$. Under an infinitesimal transformation $\delta X^{a} (\sigma^{\mu} ) = (\epsilon \cdot X )^{a} (\sigma^{\mu} ) $, where $\epsilon (\sigma^{\mu} )$ is the parameter of the transformations and the dot denotes the appropriate product between $\epsilon$ and $X^a$, the action varies as follows
\begin{equation}
\delta S^{(\alpha )}_0 [X] = \int_{-\infty}^{+\infty}  d \tau 
\int_{0}^{+\infty} d \sigma \, \partial_{\mu} \left( \epsilon \right) \cdot j^{\mu} 
\, ,
\label{fsm-fnp-symm-2}
\end{equation}
where, $j^{\mu}$ is the current associated to the transformation $\delta X^{a} (\sigma^{\mu} )$. 

Let's begin with the translation symmetry, characterized by the parameter $\xi^{a} (\sigma^{\mu} )$. It's straightforward to observe that
\begin{equation}
\delta S^{(\alpha )}_0 [X] =  \kappa_{\alpha}
\int_{-\infty}^{+\infty}  d \tau 
\int_{0}^{+\infty} d \sigma \sigma^{1-\alpha} 
\partial_{\mu} \left( \xi^{a} \right) \partial^{\mu} X_{a} 
\, .
\label{fsm-fnp-symm-3}
\end{equation}
By comparing equations (\ref{fsm-fnp-symm-2}) and (\ref{fsm-fnp-symm-3}), we deduce that the current corresponding to the translation symmetry is
\begin{equation}
j^{\mu}_{a} =  \kappa_{\alpha} \sigma^{1-\alpha} 
\partial^{\mu} X_{a} 
\, .
\label{fsm-fnp-symm-4}
\end{equation}
The current $j^{\mu}_{a}$ is conserved on-shell
\begin{equation}
\partial_{\mu} j^{\mu}_{a} =  \kappa_{\alpha} 
\partial_{\mu} \left( \sigma^{1-\alpha} 
\partial^{\mu} X_{a} \right) = 0
\, .
\label{fsm-fnp-symm-5}
\end{equation}
The corresponding momentum $p^{a} $ is given by
\begin{equation}
p^{a} (\tau ) = \int_{0}^{+\infty} d \sigma \, j^{a 0} 
= -  \kappa_{\alpha}
\int_{0}^{+\infty} d \sigma \sigma^{1-\alpha} 
\partial_{\tau} X_{a} (\tau , \sigma ) = \int_{0}^{+\infty} d \sigma P^{a} (\tau , \sigma )
\, ,
\label{fsm-fnp-symm-6}
\end{equation}
where $P_{a} (\tau , \sigma )$ is the canonical momentum conjugate to the field $X^{a} (\tau , \sigma )$, calculated at a fixed time value $\tau = \tau_0$ and defined as usual
\begin{equation}
P_{a}(\tau , \sigma ) = \frac{\partial \mathcal{L}^{(\alpha)}_{0}\left(X,\partial X\right)}{\partial \left[ \partial_{\tau} X^{a} (\tau , \sigma ) \right]}
= -  \kappa_{\alpha} \sigma^{1-\alpha} 
\partial_{\tau} X_{a} (\tau , \sigma )
\, ,
\label{fsm-fnp-mom}
\end{equation}
where $\mathcal{L}^{(\alpha)}_{0}\left(X,\partial X\right)$ is the Lagrangian density from $S^{(\alpha )}_0 [X]$.

Now, let's examine the Lorentz transformation with the infinitesimal parameter 
${\epsilon^{a}}_{b} (\sigma^{\mu} )$. The variation of the action under these transformations is
\begin{equation}
\delta S^{(\alpha )}_0 [X] =  \kappa_{\alpha}
\int_{-\infty}^{+\infty}  d \tau 
\int_{0}^{+\infty} d \sigma \sigma^{1-\alpha} 
\partial_{\mu} \left( \epsilon_{ab} \right) X^{b} \partial^{\mu} X^{a}
\, . 
\label{fsm-fnp-symm-7}
\end{equation}
From equation (\ref{fsm-fnp-symm-7}), we can see that the current corresponding to the Lorentz symmetry is 
\begin{equation}
j^{ab}_{\mu} =  \kappa_{\alpha} \left( X^{a} \partial_{\mu} X^{b} - X^{b} \partial_{\mu} X^{a}\right)
\, .
\label{fsm-fnp-symm-8}
\end{equation}
It is easy to verify that the above current is conserved on-shell
\begin{equation}
\partial^{\mu} j^{ab}_{\mu} = 0.
\label{fsm-fnp-symm-9}
\end{equation}
The current $j^{ab}_{\mu}$ is constructed to be antisymmetric due to the antisymmetry of the parameters $\epsilon_{ab}$.

From the preceding analysis, we can see that the phase space of the sigma model is parametrized by $X^{a} (\tau , \sigma )$ and $P_{a} (\tau , \sigma )$. 
Since $\tau$ is interpreted as a time-variable, we can introduce the standard Poisson structure in terms of the Poisson brackets at a fixed time
\begin{align}
& \left\{
X^{a} (\tau, \sigma) , X^{b} \left(\tau, \sigma^{\prime} \right)
\right\}_{\text{P.B.}}
=
\left\{
\partial_{\tau} X_{a}(\tau, \sigma) , \partial_{\tau} X_{b} \left(\tau, \sigma^{\prime}\right)\right\}_{\text{P.B.}}=0
\, ,
\label{fsm-fnp-Poisson-1}
\\
& 
\left\{
X^{a} \left(\tau, \sigma \right) 
,
\partial_{\tau} X_{b}(\tau, \sigma^{\prime} ) 
\right\}_{\text {P.B. }}
= \frac{\sigma^{\alpha - 1}}{\kappa_{\alpha}}
\delta^{a}_{b} \delta\left(\sigma-\sigma^{\prime}
\right) 
\, .
\label{fsm-fnp-Poisson-2}
\end{align}
This analysis reveals that the global symmetries of the sigma model derived from the Caffarelli-Silvestre procedure mirror those of the Polyakov action. While the Lagrangian densities $\mathcal{L}^{\alpha}_{0}$ and $\mathcal{L}_{\text{Polyakov}}$ have similar structures from the perspective of target space, they exhibit important distinctions when local symmetries are considered. Unlike the reparametrization invariance inherent in string theory's Polyakov action, which originates from the interpretation of the world-sheet as the surface swept out by string during its evolution,
$\mathcal{L}^{\alpha}_{0}$ explicitly depends on $\sigma$, rendering it non-reparametrization invariant. This arises due to constraints imposed by the Caffarelli-Silvestre extension problem, which dictates the form of the sigma model as a two-dimensional field theory devoid of reparametrization and Weyl symmetries, and with no direct interpretation in terms of one-dimensional physical objects. Consequently, in this context, the term 'world-sheet' serves as an extension of the string terminology, albeit without preserving the same content.

This lack of reparametrization invariance leads to notable differences, particularly in the structure of canonical momenta. In the context of bosonic string theory, two momenta densities, $ P^{\tau}_{a} $ and $ P ^{\sigma}_{a} $, govern the evolution of string coordinates along the time and space world sheet variables. Reparametrization invariance of the Lagrangian density under shifts in $ \tau $ and $ \sigma $ coordinates dictates the appearance of two momenta, each tied to one of the world sheet coordinates. However, as the sigma model derived from the extension problem lacks this reparametrization invariance, the corresponding momentum $ P ^{\sigma}_{a} $ assumes a form that lacks a straightforward physical interpretation
\begin{equation}
P^{\sigma}_{a}(\tau , \sigma ) = \frac{\partial \mathcal{L}^{(\alpha)}_{0}\left(X,\partial X\right)}{\partial \left[ \partial_{\sigma} X^{a} (\tau , \sigma ) \right]} = \kappa_{\alpha} \sigma^{1-\alpha} \partial_{\sigma} X_{a} (\tau , \sigma ).
\label{fsm-fnp-mom-1}
\end{equation}
This highlights a distinctive characteristic of the sigma model derived from the extension problem, where the canonical momenta exhibit behavior that may not align with conventional expectations.

%-----------------------------------------------------------------
%-----------------------------------------------------------------
\subsection{Canonical Quantization of Sigma Model}
\label{sec:fsm-free-quant}

Quantizing the fractional particle model directly, as discussed in previous sections, poses significant challenges due to the absence of an apparent canonical structure, even classically. However, the Caffarelli-Silvestre extension problem offers a promising avenue by enabling the extension of the fractional particle model residing on the world-sheet boundary to a local sigma model within the bulk. By quantizing this sigma model, valuable insights into both systems can be gleaned. Given the canonical structure inherent in the phase space of the sigma model, canonical quantization emerges as a natural approach. Unlike the conventional treatment of the standard sigma model, the momentum \( P ^{\sigma}_{a} \) does not hold special significance, as discussed previously, owing to its lack of interpretability as associated with reparametrization in the \( \sigma \)-direction.

The canonical quantization of the free scalar field with the Caffarelli-Silvestre extension problem was delineated in \cite{Frassino:2019yip}. Here, we follow analogous steps to quantize the sigma model. According to canonical quantization principles, the coefficients $ A^b(\omega) \rightarrow \mathrm{a}^b (\omega) $ and $ \bar{A}^b(\omega) \rightarrow \mathrm{a}^{b \dagger} (\omega) $ are elevated to operators on the Fock space of the quantum sigma model, while $ X^{a} (\tau , \sigma) $ and $ P_{a} (\tau , \sigma) $ are treated as field operators, subject to standard equal-time commutation relations
\begin{align}
\left[
X^b(\tau, \sigma)
, 
P_{c}  (\tau , \sigma^{\prime} )
\right]
& = i \delta^{b}_{c}  
\delta \left( \sigma - \sigma^{\prime} \right)
\, ,
\label{fsm-fnp-comm-rel}
\\
\left[
X^b(\tau, \sigma)
, 
X^{c} (\tau , \sigma^{\prime} )
\right]
& =
\left[
P_{b} (\tau, \sigma)
, 
P_{c}  (\tau , \sigma^{\prime} )
\right] = 0
\, .
\label{fsm-fnp-comm-rel-0}
\end{align} 
Since the equation (\ref{fsm-fnp-mom}) is invertible, the canonical Hamiltonian density $\mathcal{H}^{\alpha}_{0}$ is defined by the Legendre transformation of the Lagrangian density
\begin{align}
\mathcal{H}^{\alpha}_{0} (\tau, \sigma) & = 
P_{a}  (\tau, \sigma) \partial_{\tau} X^{a} (\tau, \sigma) - 
\mathcal{L}^{\alpha}_{0} (\tau, \sigma)
\nonumber
\\
& = - \frac{\sigma^{\alpha - 1}}{2 \kappa_{\alpha}}
\left[
P_{a}  (\tau, \sigma) P^{a} (\tau, \sigma) 
+ 
m^{2}_{\alpha} \kappa^{2}_{\alpha} \sigma^{2 \alpha - 2}
\partial_{\sigma} X_{a} (\tau, \sigma) \partial_{\sigma} X^{a} (\tau, \sigma)
\right]
\, .  
\label{fsm-fnp-hamiltonian}
\end{align}
One can easily verify the following commutation relations
\begin{align}
\left[
X^{b} (\tau , \sigma )
,
\mathcal{H}^{\alpha}_{0} (\tau , \sigma^{\prime} )
\right]
& = 
- i \frac{\sigma^{\alpha - 1}}{\kappa_{\alpha}}
\delta^{b}_{c} \, P^{c} (\tau , \sigma^{\prime}) \delta (\sigma - \sigma^{\prime} ) 
\, ,
\label{fsm-fnp-ham-comm-1}
\\
\left[
P_{b}  (\tau , \sigma )
,
\mathcal{H}^{\alpha}_{0} (\tau , \sigma^{\prime} )
\right]
& = 
i \sigma^{1-\alpha} \kappa_{\alpha}
\delta_{b c} \, \partial_{\sigma^{\prime}} X^{c} (\tau , \sigma^{\prime}) \frac{\partial \delta (\sigma - \sigma^{\prime} )}{\partial \sigma^{\prime}} 
\, .
\label{fsm-fnp-ham-comm-2}
\end{align} 
The decomposition of quantum fields into modes is given by the following relations
\begin{align}
X^b(\tau, \sigma) = 
\int_0^{+\infty} \frac{d \mu(\omega)}{\sqrt{2 \omega }}
\left[ 
\mathrm{a}^{b}(\omega) e^{-i\omega \tau } \varphi(\sigma, \omega)
+ \mathrm{a}^{b \dagger}(\omega) 
e^{i\omega \tau } \bar{\varphi}(\sigma, \omega)\right]
\, ,
\label{fsm-fnp-x-modes}
\\
P_{b}  (\tau , \sigma ) = i
\int_0^{+\infty} d \mu(\omega) \sqrt{ \frac{\omega}{2}}
\left[ 
- \mathrm{a}_{b}(\omega) e^{-i\omega \tau } \varphi(\sigma, \omega)
+ \mathrm{a}_{b}^{\dagger}(\omega) 
e^{i\omega \tau } \bar{\varphi}(\sigma, \omega)\right]
\, .
\label{fsm-fnp-momenta-modes}
\end{align}
The commutation relations among the mode operators can be derived using equations (\ref{fsm-fnp-eom-ort}), (\ref{fsm-fnp-ham-comm-1}), and (\ref{fsm-fnp-ham-comm-2}), yielding the following outcome 
\begin{align}
\left[ 
\mathrm{a}^b (\omega) ,
\mathrm{a}^c (\omega^{\prime})
\right]
& = 
\left[ 
\mathrm{a}^{b \dagger} ( \omega) ,
\mathrm{a}^{c \dagger} (\omega^{\prime})
\right]=0
\, ,
\label{fsm-fnp-comm-a1}
\\
\left[
\mathrm{a}^{b} (\omega) ,
\mathrm{a}^{c \dagger} (\omega^{\prime})
\right]
& =
\frac{2 \pi}{\mu (\omega )} \delta^{b c}
\delta\left( \omega -\omega^{\prime} \right)
\, .
\label{fsm-fnp-comm-a2}
\end{align}
From the aforementioned relations, we can see that the operators $\mathrm{a}^{b \dagger} (k, \omega)$ and $\mathrm{a}^b (k, \omega)$ create and annihilate excitations along $b$-direction on the half-plane world-sheet of the sigma model from the state $\vert 0 \rangle^{b}_{k, \omega} $, and the vacuum state is the product of all single mode vacua. The Hamiltonian has the following form
\begin{align}
H^{\alpha}_{0} = \int_{-\infty}^{+\infty} d \tau \int_{0}^{+\infty} d \sigma \, \mathcal{H}^{\alpha}_{0} (\tau, \sigma)
& = \varepsilon_{\alpha} \int_{0}^{+\infty} d \sigma
\int_{0}^{+ \infty} d \mu( \omega )
\left[
\mathrm{a}^{\dagger}_{b} (\omega) \mathrm{a}^{b} (\omega ) +
\frac{n \pi}{\mu (\omega )}
\right]
\nonumber
\\
& \times
\left[
\sigma^{\alpha - 1} \omega^{1-\alpha} 
\vert \varphi (\omega , \sigma) \vert^2
+ \kappa_{\alpha}^{2}
\sigma^{1-\alpha} \omega^{-\alpha} 
\vert \partial_{\sigma} \varphi (\omega , \sigma)\vert^2
\right]
\, , 
\label{fsm-fnp-ham-decomp}
\end{align}
where $\varepsilon_{\alpha} $ is a shorthand notation for the parameter
\begin{equation}
\varepsilon_{\alpha}= \frac{2}{\pi} \sin^{2} \left( \frac{\pi \alpha}{2}\right) \Gamma \left( 1- \frac{\alpha}{2}\right) \Gamma \left( \frac{\alpha}{2} \right)
\, .
\label{fsm-fnp-ham-epsilon}
\end{equation}
The integrals from the right hand side of the equation (\ref{fsm-fnp-ham-epsilon}) diverge. This is a common problem of the scalar field theory, which can be addressed by introducing cutoff parameters $\lambda_c$ in $\sigma$ and $\omega_c$ in $\omega$ variables \cite{Frassino:2019yip}. However, calculating the energy of the sigma model for any given state necessitates the use of computational methods.

An important consideration lies in describing the energy of the sigma model near its boundary. To derive the boundary Hamiltonian, we substitute the boundary conditions provided in equation (\ref{fsm-fnp-eom-bc}) into equation (\ref{fsm-fnp-ham-decomp}). The localization process is implemented by introducing the boundary cutoff distance $\lambda_B$, corresponding to an energy cutoff $\Lambda_B = \lambda_B^{-1}$ above which energies are integrated out. Denoting the frequency cutoff on the boundary as $\omega_B$, the boundary Hamiltonian is expressed as follows
\begin{equation}
H^{\alpha}_{0} (\lambda_B , \omega_B ) =
\varepsilon_{\alpha}
\int_{0}^{\lambda_B} d \sigma
\int_{0}^{\omega_B} d \mu( \omega )
\left[
\mathrm{a}^{\dagger}_{b} (\omega) \mathrm{a}^{b} (\omega ) +
\frac{n \pi}{\mu (\omega )}
\right]
\sigma^{\alpha - 1} \omega^{1-\alpha} 
\, . 
\label{fsm-fnp-ham-decomp-bound}
\end{equation}
The boundary vacuum energy can be calculated from the equation (\ref{fsm-fnp-ham-decomp-bound}) and it is given by
\begin{equation}
E^{\alpha}_{\mathrm{vac}} (\lambda_B, \omega_B ) = 
\langle 0 \vert H^{\alpha}_{0} (\lambda_B , \omega_B ) \vert 0 \rangle 
=
\frac{\varepsilon_{\alpha} \pi n \lambda_B^{\alpha} \omega_{B}^{2-\alpha}}{\alpha (2-\alpha)} 
\, .
\label{fsm-fnp-vac-B}
\end{equation}
If we fix the frequecy cutoff at the scale of the localization distance
$\omega_B=\Lambda_B$, the vacuum energy near the boundary behaves like $ E^{\alpha}_{\mathrm{vac}} (\Lambda_B ) = E(\alpha , \Lambda_B )\sim \Lambda_B^{2-2\alpha}$. 

In order to illustrate the boundary vacuum energy as a function of fractionality parameter $\alpha$, we give in the  Table \ref{table-vacuum-values} the formulas of $E^{\alpha}_{\mathrm{vac}} (\lambda_B, \omega_B )$ calculated from the equation (\ref{fsm-fnp-vac-B}) for several values of $\alpha$ with the assistance of the computer. In particular, $E^{0}_{\mathrm{vac}} (\lambda_B, \omega_B )$ and $E^{2}_{\mathrm{vac}} (\lambda_B, \omega_B )$ were obtained by taking the corresponding limits. For $\omega_B=\Lambda_B$ and $n=4$, we plot the boundary vacuum energy $E(\alpha, \Lambda_B )$ as a function of $\alpha$ for different values of $\Lambda_B$ in the Figure 2 from Appendix B. The family of plots is symmetric with respect to  $E(\alpha, \Lambda_B =1 ) $. If $\Lambda_B < 1$, the boundary vacuum energies have maxima for $\alpha \in (1,2)$. If $\Lambda_B > 1$, the boundary vacuum energies display their maxima for values of $\alpha \in (0,1)$. This shows that the vacuum energy of a fixed fractionality field near the boundary varies according to the cutoff scale, or, equivalently, to the distance set for localization with respect to the boundary. As $\sigma$ tends to zero, the boundary vacuum energy tends to zero, too. We interpret this result as a statement on the limitation of the quantum sigma model constructed from the extension problem to provide information on the fractional particle. Indeed, even if the classical action of the particle model and the sigma model are equal, the later is non-fractional and can be quantized while the former does not display a proper canonical structure and, consequently, cannot be quantized by using standard methods. This open up the problem of the interpretation of the fractional particle model and of its mathematical structure in a way compatible with the quantum mechanics.

%-----------------------------------------------------------------
\section{Discussions}

In the present paper, we examined the lowest-dimensional realization of the extension problem for the fractional Laplacian in terms of the classical fractional particle model in $n$-dimensions. We have discussed the equations of motion and the Green's function method, the symmetries of the free fractional particle, as well as some examples of local interactions such as the classical harmonic oscillator and the fractional charged particle in the electromagnetic field. For the free fractional particle, we have constructed the local sigma model provided by the extension problem. The fractional particle model proposed here represents a generalization of the Newtonian particle with the fractional Laplacian, while the sigma model generalizes the standard sigma model on the superior half plane. These two models generalize the application of the extension problem to the classical scalar field \cite{Frassino:2019yip}. We have quantized the sigma model by the canonical quantization method and have discussed the vacuum energy near the boundary. There are some aspects of the fractional particle and sigma model worth discussing. 

Starting with sigma model, we observe that the choice of measure $\mu (\omega )$ given in the equation (\ref{fsm-fnp-eom-mu}) is not unique. This point has been made clear in \cite{Mintchev:2001yz}, where it was emphasized that the measure adopted here is useful to describe scale invariance in higher dimensional field theories. The measure plays an important role in the quantization of the fields in the bulk \cite{Frassino:2019yip} since it determines the classical Poisson brackets which are then elevated to equal-time commutation relations between the fields $X^{a}(\tau , \sigma )$ and their canonically conjugate momentum $P_{a} (\tau , \sigma )$. An interesting problem is to explore the quantization with different measures. Another important point is the observation that the analogy with the Polyakov action raises the question of the second momentum $P^{\sigma}_{a} (\tau , \sigma )$ defined by the equation (\ref{fsm-fnp-mom-1}). Since the sigma model is not reparametrization invariant with respect to $\sigma$ variable as discussed at the end of subsection (\ref{sec:fsm-symm}), we did not use the momentum $P^{\sigma}_{a} (\tau , \sigma )$ to define the physical states. However, it is an interesting problem to make an extensive analysis of the mathematical structure and physical properties of the sigma model presented here. 

An important property of the sigma model, common to higher dimensional scalar fields, is the following. As already observed in \cite{Mintchev:2001yz,Frassino:2019yip}, the  scalar fields are non-local for a general 
$\mu (\omega )$ measure as a consequence of the defect introduced in the world-sheet by the presence of the boundary. The non-locality is a quantum effect since $\mu (\omega )$ determines the properties of the creation and annihilation operators according to the equations (\ref{fsm-fnp-comm-a1}) and (\ref{fsm-fnp-comm-a2}). In particular, this introduces a factor of $\omega^{1-\alpha}$ that multiplies the number operators, as can be seen from the equation (\ref{fsm-fnp-ham-decomp}). 
That raises the question of the emergence of oscillating modes instead of decaying/exploding modes. This is primarily influenced by the bounded domain of the problem and the specific mathematical framework utilized for the quantization process. The bounded domain of the problem, as considered in the extension problem framework, plays a significant role in shaping the behavior of field modes, since the imposition of boundary conditions within a finite domain can lead to the confinement of solutions and the promotion of oscillatory behavior favoring oscillating modes over decay or explosion. On the other hand, the extension problem by Caffarelli and Silvestre, allows one to control the behavior of field modes, and guide the system towards oscillatory solutions. Thus, we can conclude that in bounded spaces, the interplay between the non-local dynamics introduced
by the fractional Laplacian operator and the constraints imposed by the boundaries can lead to stable oscillatory modes. The confinement of the field within a finite region, coupled with the mathematical treatment of the problem, supports the existence of oscillations as viable and physically meaningful solutions.

Turning to the fractional particle model, we observe that the fractional Laplacian, characterized by its pseudo-differentiability and non-locality, presents interpretative challenges due to its inherent complexity. It is commonly interpreted in several ways: as a representation of the quantity of particles capable of transitioning between points in diffusion theory \cite{Chen:2020}; as a generalized Dirichlet-to-Neumann operator, linked with GJMS operators on smooth metric spaces \cite{Case:2016}; or as a fractional gradient, which correlates with the uniform isotropic measure \cite{Dovidio:2013}. It is crucial to underscore that these interpretations are abstract, and the actual behavior of functions under the influence of the fractional Laplacian can exhibit significant complexity, contingent on the specific problem or application. The task of attributing a geometric interpretation to the fractional operators in fractional calculus - a field replete with diverse definitions and interpretations — poses a formidable challenge. As of now, the scientific community has not arrived at a consensus regarding the geometric significance of these operators. For an in-depth discussion of this subject within the context of fractional derivatives, refer to \cite{Tarasov:2016}. 

It's important to recognize that the generalization of the standard Lagrangian in equation (\ref{fsm-fnp-act-newt}) to a fractional particle is not unique. We could have directly replaced the derivatives in $\tau$ with first-order fractional derivatives $(-\Delta^{(1)}_{\tau})^{\frac{\alpha}{2}}$ as defined in equation (\ref{fsm-fnp-L-der-1}), instead of integrating by parts. Although this is a feasible alternative, we choose the fractional Laplacian in the kinetic term. This decision is motivated by the well-established properties of $(-\Delta_{\tau})^{\frac{\alpha}{2}}$, which simplify the interpretation of the kinetic term. Additionally, defining the non-local derivative as a fractional Laplacian maintains a simpler mathematical interpretation, which is absent in $(-\Delta^{(1)}_{\tau})^{\frac{\alpha}{2}}$. Another reason for using the fractional Laplacian in the kinetic term is its equivalence, under the extension problem by Caffarelli and Silvestre \cite{Caffarelli:2007}, to the local sigma model in two dimensions discussed here \cite{Frassino:2019yip}. In a formulation based on the first fractional derivatives, we can only apply this mapping once we have integrated one of the primary fractional derivatives. This results in a Lagrangian that takes the form specified in equation (\ref{fsm-fnl-act-frac-newt-1}).

The fractional particle models and the sigma model discussed here should be further investigated in several directions. One important problem is finding solutions to the equations of motion of the fractional particle with different boundary conditions and with interaction terms, which is a non-trivial mathematical problem. That could help to better understand the correspondence between the dynamics of the fractional particle and that of the sigma model. Another interesting question is how the interactions in the fractional particle model can be mapped into the interactions of the sigma model at the classical level. While the Caffarelli-Silvestre extension problem provides a rigorous mapping of the kinetic terms between the two models, the interactions are subject to interpretation. An interesting line of inquiry concerns the construction of the Hamiltonian formalism on the two sides of the extension problem. For example, it is known that the Hamiltonian formalism is not well defined in the case of fractional Laplacian fields, as illustrated by our discussion of the linear momentum, too. On the other hand, the local sigma model Hamiltonian can be defined in the standard way. That shows a limitation in the applicability of the extension problem to the Hamiltonian formalism, which urges further analysis. 

An important problem to be investigated is related to the generalization of the present model to other domains that $\mathcal{X}$. For particles moving along finite time intervals, one can still consider the coordinates as functions from $\mathscr{C}_{b u}$. However, a detailed analysis of the boundary conditions is necessary. Besides, the relation between different definitions of the fractional Laplacian should be treated with care. For example, in \cite{Barci:1995ad,Barci:1996br,Barci:1996ny,Barci:1998wp}, fractional operators admit a Fourier-like transform only in the sense of ultradistributions, where the integral domain is on a complex contour instead of on the real line. A generalization of the particle 
model along the line of these works is interesting, since it can lead to a direct quantization of the particle model by developing the fractional Lagrangian in a power series and can produce asymptotic quantum states.

An interesting aspect of the classical fractional particle model and sigma model discussed in this work that deserves further study is the correspondence between their symmetries and their relation to the conservation of non-local quantities. A related problem is whether the fractional particle could realize the conformal algebra similarly to the higher order particle proposed in \cite{Gomis:2011dw}. And finally, we observe that it is necessary to study deeper the quantum structure of the sigma model, and in particular the role played by symmetries at different values of the fractionality parameter. We end by noting that, although the study of the fractional models discussed here is interesting from a pure mathematical and mathematical physics point of view, describing concrete physical systems with similar properties to the classical fractional particles and their corresponding sigma models is a major challenge.

%-----------------------------------------------------------------
\section*{Acknowledgments}

I acknowledge R. S. Facundo, C. F. L. Godinho and M. C. Rodriguez for discussions and J. Gomis for correspondence. I also acknowledge an anonymous referee for very constructive and important feedback which have greatly improved the quality of the present work. This work received partial support from the Basic Research Grant (APQ1) from the Carlos Chagas Filho Foundation for Research Support of the State of Rio de Janeiro (FAPERJ), grant number E-26/210.511/2024.

%-----------------------------------------------------------------
\section*{Appendix A: Basic Properties of Fractional Laplacian}
\renewcommand{\theequation}{A.\arabic{equation}} 

In this Appendix, we briefly review some basic properties of the fractional Laplacian. We refer to \cite{Kwasnicki:2017} and \cite{Chen:2020} for more details and proofs.

The fractional Laplacian, denoted as $(-\Delta_{\tau})^{\frac{\alpha}{2}}$, is a pseudo-differential operator that generalizes the concept of spatial derivatives from the Laplacian to fractional orders. The literature presents a multitude of definitions for the fractional Laplacian, equivalent with each other on certain spaces of functions, yet offering unique perspectives. In this paper, we adopt three definitions that are equivalent to each other on any of the spaces $\mathscr{L}^p, p \in[1, \infty)$, $\mathscr{C}_0$ or $\mathscr{C}_{b u}$ according to the Theorem 1.1 from \cite{Kwasnicki:2017}. Here,
$\alpha \in (0, 2)$, $p \in [1, \frac{1}{\alpha})$, $\mathscr{L}^p$ denotes the Lebesgue spaces, $\mathscr{C}_0$ denotes the space of the continuous functions and  $\mathscr{C}_{b u}$ is the space of bounded uniformly
continuous functions. Our formulas are given for functions on $\mathbb{R}$, but they are the same for functions on $\mathbb{R}^n$ with $n \in \mathbb{N}$. The definitions below are also equivalent on the Schwartz space $\mathscr{S}$ on the set of rapidly decreasing functions from $C^{\infty} (\mathbb{R})$ defined as
\begin{equation}
\mathscr{S} = \left\{ f \in C^{\infty} \left( \mathbb{R} \right) : 
\sup_{\tau \in \mathbb{R}} \left| \tau^q  f^{(p)}(\tau ) \right| < \infty, \forall \, p, q \in \mathbb{N}_0 \right\}
\, .
\label{fsm-fnp-fL-1}
\end{equation}

The first definition is given as a Fourier transform 
\begin{equation}
(-\Delta_{\tau})^{\frac{\alpha}{2}} f(\tau) = \int_{-\infty}^{+\infty} \frac{d \omega}{2 \pi } \vert \omega \vert^\alpha \tilde{f}(\omega) e^{i \omega \tau} =
\mathcal{F}^{-1} \left\{ |\omega |^{\alpha} \mathcal{F}\{f \} (\omega ) \right\} (\tau)
\, ,
\label{fsm-fnp-L-frac-def}
\end{equation}
for all $\omega \in \mathbb{R}$. Here, $\mathcal{F}$ and $\mathcal{F}^{-1}$ denote the Fourier transform and its inverse for which we use the conventions
\begin{align}
\tilde{f} (\omega) & 
:= \mathcal{F} \left\{ f(\tau ) \right\} (\omega) 
= 
\int_{- \infty}^{+\infty} d \tau f (\tau ) e^{-i \omega \tau}
\, ,
\label{fsm-fnp-L-frac-Fou}
\\
f (\tau) & 
:= \mathcal{F}^{-1} \left\{ \tilde{f}(\omega ) \right\} (\tau) 
= 
\int_{- \infty}^{+\infty} \frac{d \omega}{2 \pi} \tilde{f} (\omega ) e^{i \omega \tau}
\, .
\label{fsm-fnp-L-frac-Fou-inv}
\end{align}

The second definition is based on the interpretation of 
$(-\Delta_{\tau})^{\frac{\alpha}{2}}$ in terms of singular integrals
\begin{equation}
(-\Delta_{\tau})^{\frac{\alpha}{2}} f(\tau) = 
C_{1,\alpha} \mathrm{P. V.}\int_{-\infty}^{+\infty} d \zeta \, \frac{f(\tau ) - f(\zeta )}{\vert \tau - \zeta \vert^{1+ \alpha}} 
\, ,
\qquad
C_{1,\alpha} = \frac{2^{\alpha} \Gamma \left(\frac{1+\alpha}{2} \right)}{ \pi^{\frac{1}{2}} \vert \Gamma \left( - \frac{\alpha}{2} \right) \vert}
\, ,
\label{fsm-fnp-L-frac-def-1}
\end{equation}
for all $f \in \mathscr{S}$. The principal value operation $\mathrm{P.V.}$ is defined by the corresponding integral over the punctured one-dimensional disk $I_{\xi} (\tau)=(\tau_1 , \tau - \xi ) \cup (\tau + \xi , \tau_2)$ centred in $\tau$ and of radius $\xi>0$. The principal value is introduced to guarantee the existence of the inverse of the Fourier symbol $|\sigma|^{\alpha}$ for $\alpha \in (0, 2]$, and the continuation of $(-\Delta_{\tau})^{\frac{\alpha}{2}}$ from negative to positive values of $\alpha$.

The third definition regards $(-\Delta_{\tau})^{\frac{\alpha}{2}}$ as a distribution. For any $f \in \mathscr{S}$, we define 
\begin{equation}
\left\langle(-\Delta_{\tau})^{\frac{\alpha}{2}} f, \varphi \right\rangle
=
\int_{\mathbb{R}} d \tau \varphi(\tau) (-\Delta_{\tau})^{\frac{\alpha}{2}} f(\tau)   
=
\int_{\mathbb{R}} d \tau f(\tau)(-\Delta_{\tau})^{\frac{\alpha}{2}} \varphi(\tau)  
\, ,
\label{fsm-fnp-L-frac-def-2}
\end{equation}
for all test functions $ \varphi \in C_0^{\infty}\left(\mathbb{R}\right) 
$. In the equation (\ref{fsm-fnp-L-frac-def-2}), the factor $(-\Delta_{\tau})^{\frac{\alpha}{2}} f(\tau)$ can be defined as in the equation (\ref{fsm-fnp-L-frac-def}). 
It is important to observe that the singular integral and the distribution definitions also hold on the space $L^{1} (\mathbb{R})$ obtained by relaxing the regularity and infinity conditions on the functions
\begin{equation}
L_{\alpha}^1 \left(\mathbb{R} \right)
=
\left\{f \in L_{\mathrm{loc}}^1 \left(\mathbb{R} \right) : 
\int_{\mathbb{R}} \frac{|f(\tau)|}{1+|\tau|^{1+ \alpha}} d \tau 
< + \infty \right \}
\, .
\label{fsm-fnp-L-frac-def-0}
\end{equation} 
The definition in terms of Fourier transform can be applied to functions from $L^{1} (\mathbb{R})$, too, since $\vert \vert \tilde{f} \vert \vert \leq \vert \vert f \vert \vert $ by the triangle inequality. Thus, we can always consider functions that are locally integrable and have a suitable growth at infinity to avoid divergences. For a deeper discussion of the definitions and mathematical properties of the fractional Laplacian, see \cite{Kwasnicki:2017, Chen:2020}.

Third definition (\ref{fsm-fnp-L-frac-def-2}) elucidates that the fractional Laplacian acts as a pseudo-differential operator. This is manifested in its operation within the Fourier domain via multiplication, as opposed to differentiation within the spatial domain. By introducing the following first-order fractional operator 
\begin{equation}
\left(- \Delta^{(1)}_{\tau} \right)^{\frac{\alpha}{2}} f(\tau)  = 
c_{\alpha}^{(1)} \int_0^{\infty} \mathrm{d} \zeta \,
\frac{f(\tau+\zeta)-f(\tau-\zeta)}{\zeta^\alpha} 
\, ,
\label{fsm-fnp-L-der-1}
\end{equation}
for $0 \leq \alpha < 2$, where
\begin{equation}
c^{(1)}_{\alpha} = \frac{2^{\alpha-1}}{\sqrt{\pi}} \frac{\Gamma\left(\frac{1+\alpha}{2}\right)}{\Gamma\left(\frac{2-\alpha}{2}\right)}
\, ,
\qquad
c_{\alpha} = \alpha c^{(1)}_{\alpha}
\, ,
\label{fsm-fnp-L-der2}
\end{equation}
the fractional Laplacian can be associated to $\left(- \Delta^{(1)}_{\tau} \right)^{\frac{\alpha}{2}}$ 
through a conventional first-order derivative
\begin{equation}
\left(- \Delta_{\tau} \right)^{\frac{\alpha}{2}} \equiv
\frac{\mathrm{d}}{\mathrm{d} \tau} \left(- \Delta^{(1)}_{\tau} \right)^{\frac{\alpha}{2}}
\, .
\label{fsm-fnp-L-der-3}
\end{equation}
The details of this correspondence were given in equations (\ref{flm-L-def-reg}) through (\ref{flm-L-def-reg-2}). The operator $\left(- \Delta^{(1)}_{\tau} \right)^{\frac{\alpha}{2}}$ has the following spectral form
\begin{equation}
\left(- \Delta^{(1)}_{\tau} \right)^{\frac{\alpha}{2}} f(\tau)
=-\frac{\mathrm{i}}{\sqrt{2 \pi}} 
\int_{-\infty}^{\infty} \mathrm{d} \omega \, \omega |\omega|^{\alpha-2} 
\mathcal{F}[f](\omega) \mathrm{~e}^{- i \omega \tau} 
\, .
\label{fsm-fnp-spin-part-D1-3}
\end{equation}

The fractional Laplacian is considered equivalent to the Riesz derivative ${ }_{R Z} D_{\tau}^{\alpha}$ on $\mathbb{R}$ and on $\mathbb{R}^n$, see, for example, \cite{Chen:2020} and \cite{Cai:2019}. In the one dimensional case discussed in this work, the Riesz derivative is defined as
\begin{equation}
{ }_{R Z} D_{\tau}^{\alpha} f(\tau)=-\Psi_\alpha \left({ }_{R L} D_{-\infty, \tau}^\alpha+{ }_{R L} D_{\tau, \infty}^\alpha\right) f(\tau)
\, ,
\label{fsm-fnp-Riesz-def}
\end{equation}
where the left- and right-sided Riemann-Liouville derivatives of order $\alpha$ are defined as follows
\begin{align}
{ }_{R L} D_{-\infty, \tau}^\alpha f(\tau) & =\frac{1}{\Gamma(m-\alpha)} \frac{d^m}{d \tau^m} \int_{-\infty}^{\tau} \frac{f(\zeta)}{(\zeta-\tau)^{\alpha-m+1}} d \zeta,
\label{fsm-fnp-RL-1}
\\
{ }_{R L} D_{\tau, \infty}^\alpha f(\tau) & =\frac{(-1)^m}{\Gamma(m-\alpha)} \frac{d^m}{d \tau^m} \int_{\tau}^{\infty} \frac{f(\zeta)}{(\zeta-\tau)^{\alpha-m+1}} d \zeta
\, ,
\label{fsm-fnp-RL-2}
\end{align}
and 
\begin{equation}
\Psi_\alpha=\frac{1}{2 \cos \frac{\alpha \pi}{2}}
\, ,
\label{fsm-fnp-Riesz-1}
\end{equation}
for $\alpha \neq 1,3, \ldots$. Then, according to the Theorem 3.2 from \cite{Cai:2019}, the following relation holds
\begin{equation}
-(-\Delta_{\tau})^{\frac{\alpha}{2}} f(\tau)={ }_{R Z} D_\tau^{\alpha} f(\tau), \quad \alpha \in (0,2)/{1}
\, .
\label{fsm-fnp-Riesz-2}
\end{equation}
For more information on the properties of the fractional Laplacian, refer to \cite{Chen:2020}. For further details on the relationship between the fractional Laplacian and the Riesz derivative, consult \cite{Cai:2019}.

%-----------------------------------------------------------------
\section*{Appendix B: Boundary Vacuum Energy}
\renewcommand{\theequation}{B.\arabic{equation}} 

In this Appendix, we firstly present the boundary vacuum energy for several values of the fractionality parameter $\alpha$ obtained from the formula (\ref{fsm-fnp-vac-B}) in Table \ref{table-vacuum-values}. 

\begin{longtable}{| p{.20\textwidth} | p{.40\textwidth} |} 
\hline
$\alpha$ & $E^{\alpha}_{\mathrm{vac}} (\lambda_B, \omega_B )$ 
\\
\hline
$0$ & $\frac{n \pi^2}{2} \omega_B^2$ 
\\
\hline
$1/5$ &  $\frac{n 25 (\sqrt{5} - 1)\pi}{18} \sqrt[5]{\lambda_B \omega_B^9}$   \\
\hline
$1/4$ & $\frac{n 32 \pi}{7}\sin\left( \frac{\pi}{8}\right) \sqrt[4]{\lambda_B^3 \omega_B^5}$   
\\
\hline
$1/3$ &    
$\frac{n 3 \pi}{5} \sqrt[3]{\lambda_B \omega_B^5}$ 
\\
\hline
$2/5$ &  
$\frac{n 625 (5-\sqrt{5)}\pi}{256} \frac{1}{5}! \times \frac{4}{5}!\sqrt[5]{\lambda_B^2 \omega_B^8}$
\\
\hline
$1/2$ & $\frac{n 4 \sqrt{2}\pi}{3} \sqrt{\lambda_B \omega_B^3}$  
\\
\hline
$3/5$ &
$\frac{n 25 (1+\sqrt{5})\pi}{42} \sqrt[5]{\lambda_B^3 \omega_B^7}$
\\
\hline
$2/3$ &
$\frac{n 9 \sqrt{3}\pi}{8} \sqrt[3]{\lambda_B^2 \omega_B^4}$
\\
\hline
$3/4$ &   
$\frac{n 8 \sqrt{2}\pi}{15} \csc \left( \frac{\pi}{8} \right) 
\sqrt[4]{\lambda_B^3 \omega_B^5}$
\\
\hline
$4/5$ & 
$\frac{n 25 \sqrt{2}\pi}{24} 
\sqrt{\frac{5+\sqrt{5}}{2}}
\sqrt[5]{\lambda_B^4 \omega_B^6}$
\\
\hline
$1$ & 
$n 2\pi \lambda_B \omega_B$
\\
\hline
$4/3$ &
$\frac{n 9 \sqrt{3}\pi}{8} 
\sqrt[3]{\lambda_B^4 \omega_B^2}$
\\
\hline
$3/2$ & 
$\frac{n 4 \sqrt{2}\pi}{3} 
\sqrt{\lambda_B^3 \omega_B}$
\\
\hline
$7/4$ &
$\frac{n 32 \pi}{7} \sin \left( \frac{\pi}{8}\right) 
\sqrt[4]{\lambda_B^7 \omega_B}$
\\
\hline
$2$ &  $\frac{n \pi^2 }{2} \lambda_B^2$  \\
\hline
\caption{Boundary vacuum energy for selected values of fractionality parameter $\alpha$.}
\label{table-vacuum-values}
\end{longtable}

Next, in Figure \ref{plots-vacuum-curves} below, we plot the curves that describe the vacuum energy denoted here $E(\alpha , \Lambda_B)$ for $\omega_B = \Lambda_B$ and $n=4$. The energy scale is chosen for computational convenience and the values of $\Lambda_B$ run from 
$\textcolor{blue}{0.14} \div 7.0$. 
\begin{figure}[!htbp]
  \centering
  \includegraphics[width=12cm]{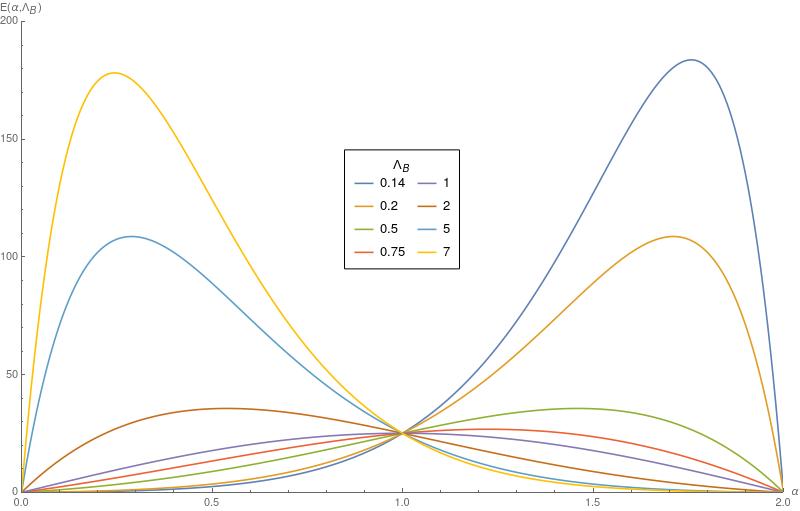}
  \caption{Boundary vacuum energy $E(\alpha, \Lambda_B )$ in $n=4$ dimensions.}
  \label{plots-vacuum-curves}
\end{figure}
For lower target space dimensions, the plots are flattened, while for high dimensions the curves are amplified. Since the vacuum energy goes as $\Lambda_B^{2-2\alpha}$, the curves form a symmetric set with respect to $E(\alpha, \Lambda_B =1)$.

\newpage

%%%%%%%%%%%%%%%%%%%%%%%%%%%%%%%%%%%%%%%%%%%%%%%%%%%%%%%%


\begin{thebibliography}{100}

%--------- NONLOCAL FIELDS ON BRANES ----------------------------

%\cite{Calcagni:2007ru}
\bibitem{Calcagni:2007ru}
G.~Calcagni, M.~Montobbio and G.~Nardelli,
	\emph{A Route to nonlocal cosmology},
	\href{https://10.1103/PhysRevD.76.126001}{Phys. Rev. D \textbf{76}, 126001
	(2007)}
	[\href{https://arxiv.org/abs/0705.3043}{{\ttfamily 0705.3043}}]. 

%\cite{Calcagni:2007ef}
\bibitem{Calcagni:2007ef}
G.~Calcagni, M.~Montobbio and G.~Nardelli,
	\emph{Localization of nonlocal theories},
	\href{https://10.1016/j.physletb.2008.03.024}{Phys. Lett. B \textbf{662}, 285-289
	(2008)}
	[\href{https://arxiv.org/abs/0712.2237}{{\ttfamily 0712.2237}}].

%\cite{Calcagni:2018lyd}
\bibitem{Calcagni:2018lyd}
G.~Calcagni, L.~Modesto and G.~Nardelli,
	\emph{Initial conditions and degrees of freedom of non-local gravity},
	\href{https://10.1007/JHEP05(2018)087}{JHEP \textbf{05}, 087 (2018),
	[erratum: JHEP \textbf{05}, 095 (2019)]}
	[\href{https://arxiv.org/abs/1803.00561}{{\ttfamily 1803.00561}}].


%--------- BOSONIC CUBIC STRING FIELD THEORY --------------------

%\cite{Witten:1985cc}
\bibitem{Witten:1985cc}
E.~Witten,
\emph{Noncommutative Geometry and String Field Theory},
	\href{https://10.1016/0550-3213(86)90155-0}{Nucl. Phys. B \textbf{268}, 253-294
	(1986)}.

%--------- LOCAL FIELDS ON BRANES NONLOCAL FIELDS IN THE BULK ---

%\cite{Mintchev:2001yz}
\bibitem{Mintchev:2001yz}
M.~Mintchev,
	\emph{Local fields on the brane induced by nonlocal fields in the 
	bulk},
	\href{https://10.1088/0264-9381/18/22/306}{Class. Quant. Grav. 
	\textbf{18},
	4801-4812 (2001)}
	[\href{https://arxiv.org/abs/hep-th/0103259}{{\ttfamily 0103259}}].

%--------- EXTENSION PROBLEM BY CAFFARELLI AND SLIVESTRE --------

%\cite{Caffarelli:2007}
\bibitem{Caffarelli:2007}
L.~Caffarelli and L.~Silvestre, 
	\emph{ An extension problem related to the fractional Laplacian}, 
	\href{https://doi.org/10.1080/03605300600987306}{Commun. Partial Differ.
	Equations \textbf{32}, 1245 (2007)}
	[\href{https://arxiv.org/abs/math/0608640}{{\ttfamily 0608640}}].

%--------- NONLOCAL QFT WITH EXTENSION PROBLEM ------------------

%\cite{Rajabpour:2011qr}
\bibitem{Rajabpour:2011qr}
M.~A.~Rajabpour,
	\emph{Conformal symmetry in non-local field theories},
	\href{doi:10.1007/JHEP06(2011)076}{JHEP \textbf{06}, 076 (2011)}
	[\href{https://arxiv.org/abs/1103.3625}{\ttfamily 1103.3625}]

%\cite{Paulos:2015jfa}
\bibitem{Paulos:2015jfa}
M.~F.~Paulos, S.~Rychkov, B.~C.~van Rees and B.~Zan,
	\emph{Conformal Invariance in the Long-Range Ising Model},
	\href{doi:10.1016/j.nuclphysb.2015.10.018}{Nucl. Phys. B \textbf{902}, 246-291
	(2016)}
	[\href{https://arxiv.org/abs/1907.007331509.00008}{{\ttfamily 1509.00008}}].

%\cite{Frassino:2019yip}
\bibitem{Frassino:2019yip}
A.~M.~Frassino and O.~Panella,
	\emph{ Quantization of nonlocal fractional field theories via the extension
	problem},
	\href{https://doi:10.1103/PhysRevD.100.116008}{Phys. Rev. D \textbf{100}, 
	no.11, 116008 (2019)}
	[\href{https://arxiv.org/abs/1907.00733}{{\ttfamily 1907.00733}}]. 

%--------- POWER SERIES DECOMPOSITION QUANTIZATION --------------

%\cite{Barci:1995ad}
\bibitem{Barci:1995ad}
D.~G.~Barci, L.~E.~Oxman and M.~Rocca,
	\emph{Canonical quantization of nonlocal field equations},
	\href{https://doi:10.1142/S0217751X96001061}{Int. J. Mod. Phys. A \textbf{11},
	2111-2126 (1996)}
	[\href{https://arXiv:hep-th/9503101}{{\ttfamily hep-th/9503101}}].

%\cite{Barci:1996br} 
\bibitem{Barci:1996br}
D.~G.~Barci, C.~G.~Bollini, L.~E.~Oxman and M.~C.~Rocca,
	\emph{Nonlocal pseudodifferential operators},
	[\href{https://arXiv:hep-th/9606183}{{\ttfamily hep-th/9606183}}].

%\cite{Barci:1996ny}
\bibitem{Barci:1996ny}
D.~G.~Barci and L.~E.~Oxman,
	\emph{Asymptotic states in nonlocal field theories},
	\href{https://doi:10.1142/S0217732397000510}{Mod. Phys. Lett. A \textbf{12},
	493-500 (1997)}
	[\href{https://arXiv:hep-th/9611147}{\ttfamily hep-th/9611147}].

%\cite{Barci:1998wp}
\bibitem{Barci:1998wp}
D.~G.~Barci, C.~G.~Bollini, M.~C.~Rocca and L.~E.~Oxman,
	\emph{Lorentz-invariant pseudo-differential wave equations},
	\href{https://doi:10.1023/A:1026696132216}{Int. J. Theor. Phys. \textbf{37},
	3015-3030 (1998)}

%--------- REVIEW NONLOCAL QFT WITH FRACTIONAL OPERATORS --------

%\cite{Calcagni:2021ljs}
\bibitem{Calcagni:2021ljs}
G.~Calcagni,
	\emph{Quantum scalar field theories with fractional operators},
	\href{https://10.1088/1361-6382/ac103c}{Class. Quant. Grav. \textbf{38}, no.16,
	165006 (2021)}
	[\href{https://arxiv.org/abs/2102.03363}{{\ttfamily 2102.03363}}]. 

%--------- NONLOCAL STRING INSPIRED FIELD THEORIES --------------

%\cite{Erbin:2021hkf}
\bibitem{Erbin:2021hkf}
H.~Erbin, A.~H.~F\i{}rat and B.~Zwiebach,
	\emph{Initial value problem in string-inspired nonlocal field theory},
	\href{https://10.1007/JHEP01(2022)167}{JHEP \textbf{01}, 167 (2022)}
	[\href{https://arxiv.org/abs/2111.03672}{{\ttfamily 2111.03672}}].

%\cite{Nortier:2021six}
\bibitem{Nortier:2021six}
F.~Nortier,
	\emph{Extra Dimensions and Fuzzy Branes in String-inspired Nonlocal Field
	Theory},
	\href{https://10.5506/APhysPolB.54.6-A2}{Acta Phys. Polon. B \textbf{54},
	no.6, 6-A2 (2023)}
	[\href{https://arxiv.org/abs/2112.15592}{{\ttfamily 2112.15592}}]

%--------- LOCAL INTERACTIONS IN NONLOCAL FIELD THEORIES --------

%\cite{Calcagni:2022shb}
\bibitem{Calcagni:2022shb}
G.~Calcagni and L.~Rachwa\l{},
	\emph{Ultraviolet-complete quantum field theories with fractional
	operators},
	\href{https://10.1088/1475-7516/2023/09/003}{JCAP \textbf{09}, 003
	(2023)}
	[\href{https://arxiv.org/abs/2210.04914}{{\ttfamily 2210.04914}}].

%--------- NONLOCAL INTERACTIONS IN QFT -------------------------

%\cite{Tomboulis:2015gfa}
\bibitem{Tomboulis:2015gfa}
E.~T.~Tomboulis,
	\emph{Nonlocal and quasilocal field theories},
	\href{doi:10.1103/PhysRevD.92.125037}{Phys. Rev. D \textbf{92}, no.12, 125037
	(2015)}
	[\href{https://arxiv.org/abs/1507.00981}{{\ttfamily 1507.00981}}].


%--------- NONLOCAL INTERACTIONS IN STRING FIELD THEORY ---------

%\cite{Pius:2016jsl}
\bibitem{Pius:2016jsl}
R.~Pius and A.~Sen,
	\emph{ Cutkosky rules for superstring field theory},
	\href{doi:10.1007/JHEP10(2016)024}{JHEP \textbf{10}, 024 (2016); erratum: JHEP
	\textbf{09}, 122 (2018)}	
	[\href{https://arxiv.org/abs/1604.01783}{{\ttfamily 1604.01783}}].	

%\cite{Pius:2018crk}
\bibitem{Pius:2018crk}
R.~Pius and A.~Sen,
	\emph{ Unitarity of the Box Diagram},
	\href{https://doi:10.1007/JHEP11(2018)094}{JHEP \textbf{11}, 094 (2018)}	
	[\href{https://arxiv.org/abs/1805.00984}{{\ttfamily 1805.00984}}].

%\cite{Chin:2018puw}
\bibitem{Chin:2018puw}
P.~Chin and E.~T.~Tomboulis,
	\emph{Nonlocal vertices and analyticity: Landau equations and general Cutkosky
	rule},
	\href{doi:10.1007/JHEP06(2018)014}{JHEP \textbf{06}, 014 (2018)}
	[\href{https://arxiv.org/abs/1803.08899}{{\ttfamily 1803.08899}}].

%--------- NON-LOCAL FREE SCALAR FIELD WITH FRACTIONAL LAPLACIAN --
%--------- APPLICATIONS -------------------------------------------

%\cite{Giusti:2020rul}
\bibitem{Giusti:2020rul}
A.~Giusti,
	\emph{ MOND-like Fractional Laplacian Theory},
	\href{https://10.1103/PhysRevD.101.124029}{Phys. Rev. D \textbf{101}, no.12,
	124029 (2020)}
	[\href{https://arxiv.org/abs/2002.07133}{{\ttfamily 2002.07133}}].
	
%\cite{Giusti:2020kcv}
\bibitem{Giusti:2020kcv}
A.~Giusti, R.~Garrappa and G.~Vachon,
	\emph{ On the Kuzmin model in fractional Newtonian gravity},
	\href{https://10.1140/epjp/s13360-020-00831-9}{Eur. Phys. J. Plus \textbf{135},
	no.10, 798 (2020)}
	[\href{https://arxiv.org/abs/2009.04335}{{\ttfamily 2009.04335}}].
	
%--------- NON-LOCAL SCALAR FIELD ENTANGLEMENT ENTROPY ------------

%\cite{Roy:2021akq}
\bibitem{Roy:2021akq}
P.~Roy,
	\emph{ Aspects of entanglement in non-local field theories with fractional
	Laplacian},
	\href{https://10.1007/JHEP06(2022)101}{JHEP \textbf{06}, 101 (2022)}
	[\href{https://arxiv.org/abs/2112.13641}{{\ttfamily 2112.13641}}].


%--------- NON-LOCAL SCALAR FIELD MASSIVE SELF-INTERACTING --------

%\cite{Basa:2019ywr}
\bibitem{Basa:2019ywr}
B.~Basa, G.~La Nave and P.~W.~Phillips,
	\emph{Classification of nonlocal actions: Area versus volume entanglement
	entropy},
	\href{https://10.1103/PhysRevD.101.106006}{Phys. Rev. D \textbf{101}, no.10,
	106006 (2020)}
	[\href{https://arxiv.org/abs/1907.09494}{{\ttfamily 1907.09494}}].

%\cite{Basa:2020cyn}
\bibitem{Basa:2020cyn}
B.~Basa, G.~La Nave and P.~W.~Phillips,
	\emph{Nonlocal Conformal Theories Have State-dependent Central Charges},
	[\href{https://arxiv.org/abs/2011.04662}{{\ttfamily 2011.04662}}].

%--------- NON-LOCAL MAXWELL CAFFARELLI ------------------------------

%\cite{LaNave:2017nex}
\bibitem{LaNave:2017nex}
G.~La Nave and P.~W.~Phillips,
	\emph{Anomalous Dimensions for Boundary Conserved Currents in Holography via the
	Caffarelli\textendash{}Silvestre Mechanism for p-forms},
	\href{https://10.1007/s00220-019-03292-z}{Commun. Math. Phys. \textbf{366}, no.1,
	119-137 (2019)}
	[\href{https://arxiv.org/abs/1708.00863}{{\ttfamily 1708.00863}}].

%\cite{LaNave:2019mwv}
\bibitem{LaNave:2019mwv}
G.~La Nave, K.~Limtragool and P.~W.~Phillips,
	\emph{Fractional Electromagnetism in Quantum Matter and High-Energy Physics},
	\href{https://10.1103/RevModPhys.91.021003}{Rev. Mod. Phys. \textbf{91}, no.2,
	021003 (2019)}
	[\href{https://arxiv.org/abs/1904.01023}{{\ttfamily 1904.01023}}].

%\cite{Porto:2023}
\bibitem{Porto:2023}
C.~M.~Porto, C.~F~.~L.~Godinho and I.~V.~Vancea,
	\emph{Fractional Laplacian Spinning Particle in External Electromagnetic
	Field},
	\href{https://doi.org/10.3390/dynamics3040046}{Dynamics \textbf{3}, 
	no. 4, 855-870 (2023)}.
	
%\cite{Heydeman:2022yni}
\bibitem{Heydeman:2022yni}
M.~Heydeman, C.~B.~Jepsen, Z.~Ji and A.~Yarom,
	\emph{Polyakov\textquoteright{}s confinement mechanism for generalized Maxwell
	theory},
	\href{https://10.1007/JHEP04(2023)119}{JHEP \textbf{04}, 119 (2023)}
	[\href{https://arxiv.org/abs/2212.11568}{{\ttfamily 2212.11568}}].

%--------- FRACTIONAL SCHROEDINGER EQUATION -----------------------

%\cite{Laskin:1999tf}
\bibitem{Laskin:1999tf}
N.~Laskin,
	\emph{Fractional quantum mechanics and Levy paths integrals},
	\href{https://10.1016/S0375-9601(00)00201-2}{Phys. Lett. A
	\textbf{268}, 298-305 (2000)}
	[\href{https://arxiv.org/abs/hep-ph/9910419}{{\ttfamily hep-ph/
	9910419}}].

%--------- NONLOCAL PARTICLES AS STRINGS --------------------------

%\cite{Cheng:2008qz}
\bibitem{Cheng:2008qz}
T.~C.~Cheng, P.~M.~Ho and T.~K.~Lee,
	\emph{Nonlocal Particles as Strings},
	\href{https:\\doi:10.1088/1751-8113/42/5/055202}{J. Phys. A
	\textbf{42}, 055202 (2009)}
	[\href{https://arxiv.org/abs/0802.1632v2}{{\ttfamily 0802.1632}}].
%4 citations counted in INSPIRE as of 01 Sep 2023

%\cite{Calcagni:2013eua}
\bibitem{Calcagni:2013eua}
G.~Calcagni and L.~Modesto,
	\emph{Nonlocality in string theory},
	\href{https://10.1088/1751-8113/47/35/355402}{J. Phys. A \textbf{47},
	 no.35, 355402 (2014)}
	[\href{https://arxiv.org/abs/1310.4957}{{\ttfamily 1310.4957}}].

%--------- POTENTIAL APPLICATIONS OF FRACTIONAL PARTICLE MODEL ----

%\cite{Bock:2020}
\bibitem{Bock:2020}
W.~Bock, J.~B.~ Bornales, C.~ O.~ Cabahug, T.~ Fattler, L.~ Streit,
	\emph{Fractional Brownian motion - Some recent results and
	generalizations}.
	In 
	\emph{9th Jagna International Workshop: Stochastic Analysis 
	– Mathematical Methods and Real-World Models, 8–18 January 2020,
	Bohol, Philippines},
	\href{https://doi.org/10.1063/5.0029699}{AIP Conf. Proc. \textbf{2286},
	020001 (2020)}.
	
%\cite{Failla:2020}
\bibitem{Failla:2020}
G.~Failla and M.~Zingales,
	\emph{Advanced materials modelling via fractional calculus: challenges
	 and perspectives},
	 \href{https://doi.org/10.1098/rsta.2020.0050}{Trans. R. Soc. A. 
	 \textbf{378}, 20200050 (2020)}.

%\cite{Zecova:2015}
\bibitem{Zecova:2015}
M.~\v{Z}ecov\'{a} and J.~Terp\'{a}k,
	\emph{Heat conduction modeling by using fractional-order derivatives},
	\href{https://www.sciencedirect.com/science/article/pii/S0096300314017974}{Applied Mathematics and Computation \textbf{257}, 365-373 2015)}

%\cite{Suzuki:2023}
\bibitem{Suzuki:2023}
J.~L.~Suzuki, M.~Gulian, M.~ Zayernouri, et al. 
	\emph{Fractional Modeling in Action: a Survey of Nonlocal Models for
	 Subsurface Transport, Turbulent Flows, and Anomalous Materials}, 
	\href{https://doi.org/10.1007/s42102-022-00085-2}{J. Peridyn. Nonlocal
	 Model \textbf{5}, 392–459 (2023)}. 

%\cite{Vasquez:2012}
\bibitem{Vasquez:2012}
J.~L.~V\'{a}zquez, 
	\emph{Nonlinear Diffusion with Fractional Laplacian Operators}, 
	In: H.~Holden, K.~Karlsen, K. (eds) 
	\emph{Nonlinear Partial Differential Equations} 
	\href{https://doi.org/10.1007/978-3-642-25361-4_15}{Abel Symposia, 
	\textbf{Vol.7} Springer, Berlin, Heidelberg, (2012)}.

%--------- FRACTIONAL POWER LAPLACIAN -----------------------------

%\cite{Kwasnicki:2017}
\bibitem{Kwasnicki:2017}
M.~Kwa\'{s}nicki,
	\emph{ Ten Equivalent Definitions of the Fractional Laplace Operator},
	\href{https://doi.org/10.1515/fca-2017-0002}{FCAA 20, 7–51 (2017)}
	[\href{https://arxiv.org/abs/1507.07356v2}{{\ttfamily 1507.07356}}]. 

%\cite{Chen:2020}
\bibitem{Chen:2020}
W.~Chen, Y.~Li and P.~Ma,
	\emph{The Fractional Laplacian},
	\href{https://doi.org/10.1142/10550}{World Scientific, (2020)}.

%---------- FRACTIONAL CALCULUS BOOKS STANDARD --------------------
%\cite{Singh:2022}
\bibitem{Singh:2022}
H.~Singh, H.~M.~Srivastava and J.~J.~Nieto, 
	\emph{Handbook of Fractional Calculus for Engineering and Science},
		\href{https://doi.org/10.1201/9781003263517}{Chapman and Hall, (2022)}.

%---------- CONFORMAL PARTICLE ------------------------------------

%\cite{Jackiw:1972cb}
%\bibitem{Jackiw:1972cb}
%R.~Jackiw,
%	\emph{Introducing scale symmetry},
%	\href{https://doi:10.1063/1.3070673}{Phys. Today \textbf{25N1}, 23-27 (1972)}

%-----------------------------


%\cite{Dyda:2020}
\bibitem{Dyda:2020}
B.~Dyda, A.~Kuznetsov, and M.~Kwaśnicki, 
	\emph{Fractional Laplace Operator and Meijer G-function},
	\href{https://doi.org/10.1007/s00365-016-9336-4}{Constr Approx 45, 427–448 (2017)}
	[\href{https://arxiv.org/abs/1509.08529}{{\ttfamily 1509.08529}}].

%---------- FRACTIONAL LAPLACE EQUATION ---------------------------

%\cite{Riesz:1938a}
\bibitem{Riesz:1938a}
M.~Riesz,
	\emph{Int\'egrales de Riemann–Liouville et potentiels},
	\href{https://acta.bibl.u-szeged.hu/13487/}{Acta Sci. Math. Szeged \textbf{9},
	1-42 (1938)}.
	
%\cite{Riesz:1938b}
\bibitem{Riesz:1938b}
M.~Riesz, 
	\emph{Rectification au travail “Intégrales de Riemann–Liouville et potentiels"},
	\href{https://acta.bibl.u-szeged.hu/13497/}{Acta Sci. Math. Szeged, \textbf{9},
	116-118 (1938)}.

%---------- RIESZ POTENTIAL ---------------------------------------
%\cite{Samko:2001}
\bibitem{Samko:2001}
S.~Samko, 
	\emph{Hypersingular Integrals and Their Applications},
	\href{https://doi.org/10.1201/9781482264968}{CRC Press, London–New York, (2001)}.

%---------- REVIEW SOLUTIONS LAPLACIAN ----------------------------

%\cite{Kwasnicki:2019}
\bibitem{Kwasnicki:2019}
M. Kwa\'snicki,
	\emph{Fractional Laplace operator and its properties},
	\href{https://doi.org/10.1515/9783110571622-007}{in A.~Kochubei and
	Y.~Luchko, Eds., \emph{Handbook of Fractional Calculus with
	Applications}, 	Volume 1 Basic Theory, De Gruyter (2019)}.	


%---------- SOLUTIONS OF FRACTIONAL LAPLACE EQ. IN TERMS OF RIESZ POTENTIAL

%\cite{Blumenthal:1961}
\bibitem{Blumenthal:1961}
R.~M.~Blumenthal, R.~Getoor and D.~Ray,
	\emph{On the distribution of first hits for the symmetric stable processes},
	\href{https://api.semanticscholar.org/CorpusID:120538923}{Trans. Am. Math. Soc.
	\textbf{99}, 540-554 (1961)}.

%---------- s-HARMONIC FUNCTIONS BOOK -----------------------------

%\cite{Landkof:1972}
\bibitem{Landkof:1972}
N~S.~Landkof,
	\emph{Foundations of Modern Potential Theory},
	\href{https://link.springer.com/book/9783642651854}{Springer Berlin, Heidelberg,
	(1972)}.

%---------- s-HARMONIC SOLUTIONS TO FRACTIONAL DIRICHLET EQ. ------

%\cite{Abatangelo:2015}
\bibitem{Abatangelo:2015}
N.~Abatangelo,
	\emph{Large s-harmonic functions and boundary blow-up solutions for the fractional
	laplacian},
	\href{https://10.3934/dcds.2015.35.5555}{Discr. Cont. Dyn. Syst. A,			\textbf{35}	(12), 5555–5607 (2015)}
	[\href{https://arxiv.org/abs/1310.3193}{{\ttfamily 1310.3193}}].

%---------- SOLUTIONS OF FRACTIONAL LAPLACE EQUATION --------------

%\cite{Hmissi:1994}
\bibitem{Hmissi:1994}
F.~Hmissi, 
	\emph{Fonctions harmoniques pour les potentiels de Riesz sur la boule unit\'{e}},
	\href{https://mathscinet.ams.org/mathscinet/relay-station?mr=1295711}{Exposition.
	Math. \textbf{12} (3), 281–288 (1994)}.

%\cite{Bogdan:1999}
\bibitem{Bogdan:1999}
K.~Bogdan,
	\emph{Representation of $\alpha$-harmonic functions in Lipschitz domains},
	\href{https://10.32917/hmj/1206125005}{Hiroshima Math. J. \textbf{29} (2), 
	227–243 (1999)}.

%---------- APPLICATIONS OF HARMONIC OSCILLATOR IN PHYSICS --------

%\cite{Stanislavsky:2004}
\bibitem{Stanislavsky:2004}
	A.~A.~Stanislavsky,
	\emph{Fractional oscillator}
	\href{https://link.aps.org/doi/10.1103/PhysRevE.70.051103}{Phys. Rev. E  \textbf{70}, 051103 (2004)}.

%\cite{Herrmann:2018}
\bibitem{Herrmann:2018}
	R.~Hermann,
	\emph{Fractional Calculus: An Introduction For Physicists},
	\href{}{World Scientific: Singapore,(2018)}.
	
%\cite{Giesel:2021wsy}
\bibitem{Giesel:2021wsy}
	K.~Giesel and A.~Vetter,
	\emph{Coherent States for Fractional Powers of the Harmonic Oscillator
	Hamiltonian},
	\href{https://10.3390/universe7110442}{Universe \textbf{7}, no.11, 442
	(2021)}
	[\href{https://arxiv.org/abs/2109.06104v1}{{\ttfamily 2109.06104v1}}].

%\cite{Quesne:2002hj}
\bibitem{Quesne:2002hj}
	C.~Quesne,
	\emph{Fractional supersymmetric quantum mechanics, topological
	invariants and generalized deformed oscillator algebras},
	\href{https://10.1142/S021773230300954X}{Mod. Phys. Lett. A 
	\textbf{18}, 515-526 (2003)}
	[\href{https://arxiv.org/0211019}{{\ttfamily 0211019}}].

%\cite{Daoud:2003qy}
\bibitem{Daoud:2003qy}
	M.~Daoud and M.~Kibler,
	\emph{Fractional supersymmetric quantum mechanics as a set of replicas
	of ordinary supersymmetric quantum mechanics},
	\href{https://10.1016/j.physleta.2003.12.027}{Phys. Lett. A
	\textbf{321}, 147-151 (2004)}
	[\href{https://arxiv.org/math-ph/0312019}{{\ttfamily 0312019}}].

%\cite{Eab:2006}
\bibitem{Eab:2006}
	C.~ H.~ Eab, S.~C.~Lim,
	\emph{Path integral representation of fractional harmonic oscillator},
	\href{https://10.1016/j.physa.2006.03.029}{Phys. A. \textbf{371},
	303-316, (2006)}.

%---------- BESSEL FUNCTIONS BOOK ---------------------------------

%\cite{Abramowitz:1972}
\bibitem{Abramowitz:1972}
M.~Abramowitz and I.~A.~Stegun,
	\emph{Handbook of Mathematical Functions with Formulas, Graphs, and
	Mathematical Tables},
	\href{}{New York: Dover, (1972)}.


%---------- INTERPRETATION OF FRACTIONAL LAPLACIAN ----------------

%\citep{Case:2016}
\bibitem{Case:2016}
J.~S.~Case and S.~Y.~Chang,
	\emph{On Fractional GJMS Operators},
	\href{https://doi.org/10.1002/cpa.21564}{Commun. Pur. Appl. Math.
	\textbf{69}, 1017- 1061 (2016)}
	[\href{https://arxiv.org/abs/1109.3773}{{\ttfamily 1109.3773}}].

%\cite{Dovidio:2013}
\bibitem{Dovidio:2013}
M.~D'Ovidio and R.~Garra,
	\emph{Fractional gradient and its application to the fractional
	 advection equation},
	[\href{https://arxiv.org/abs/1406.1846}{{\ttfamily 1406.1846}}].

%\cite{Tarasov:2016}
\bibitem{Tarasov:2016}
V.~E.~Tarasov, 
	\emph{Geometric Interpretation of Fractional-Order Derivative},
	\href{https://doi.org/10.1515/fca-2016-0062}{FCAA \textbf{19},
	 1200-1221 (2016)}.
	
%---------- CONFORMAL REALIZATION GENERALIZATION ------------------

%\cite{Gomis:2011dw}
\bibitem{Gomis:2011dw}
J.~Gomis and K.~Kamimura,
	\emph{Schrodinger Equations for Higher Order Non-relativistic Particles
	 and N-Galilean Conformal Symmetry},
	\href{https://doi:10.1103/PhysRevD.85.045023}{Phys. Rev. D \textbf{85},
	045023 (2012)}
	[\href{https://arxiv.org/abs/1109.3773}{{\ttfamily 1109.3773}}].


%---------- EQUIVALENCE FRACTIONAL LAPLACIAN AND RIESZ DERIVATIVE --

%\cite{Cai:2019}
\bibitem{Cai:2019}
M.~Cai, and C.~Li, 
	\emph{On Riesz derivative},
	\href{https://doi.org/10.1515/fca-2019-0019}{
	Frac. Calc. and App. Anal. \textbf{85}, 287-301 (2019)}. 



\end{thebibliography}
\end{document}